\documentclass[longauth]{aa}  

\usepackage{graphicx}
\usepackage{lettrine}
\usepackage[colorlinks,citecolor=blue,hypertexnames=true,bookmarks=false,breaklinks=true]{hyperref}

\usepackage{color}
\usepackage{ulem}
\usepackage{float}
\usepackage{threeparttable}
\usepackage{subdepth}

\newcommand{\sersic}{S${\rm \acute{e}}$rsic}

\newcommand{\emm}[1]{\ensuremath{#1}}
\newcommand{\emr}[1]{\ensuremath{\mathrm{#1}}}
\newcommand{\paren}[1]{\left(  #1 \right)}  
\newcommand{\bracket}[1]{\left[  #1 \right]}  

\newcommand{\spergel}[2]{\emm{\Sigma_{(\nu,#1)}^\emr{#2}}}
\newcommand{\ftspergel}[2]{\emm{\widetilde{\Sigma}_{(\nu,#1)}^\emr{#2}}}

\newcommand{\fnu}{\emm{f_{\nu}}}
\newcommand{\cnu}{\emm{c_{\nu}}}
\newcommand{\Knu}{\emm{K_{\nu}}}
\newcommand{\Lo}{\emm{L_0}}
\newcommand{\Ro}{\emm{R_0}}

\newcommand{\fisher}{\emm{\boldsymbol{I}_F}}
\newcommand{\CRB}{\emm{{\cal B}}}
\newcommand{\nvisi}{\emm{n_\emr{visi}}}
\newcommand{\visi}{\emm{V}}

\newcommand{\Rmaj}{\emm{R_\emr{maj}}}
\newcommand{\Rmin}{\emm{R_\emr{min}}}
\newcommand{\rmaj}{\emm{r_\emr{maj}}}
\newcommand{\rmin}{\emm{r_\emr{min}}}
\newcommand{\PA}{\emr{\phi}}
\newcommand{\urot}{\emm{u_\emr{rot}}}
\newcommand{\vrot}{\emm{v_\emr{rot}}}
\newcommand{\uvdist}{\emm{\rmaj^2 \urot^2 + \rmin^2 \vrot^2}}
\newcommand{\g}{\emm{g}}
\newcommand{\p}{\emm{p}}
\begin{document}
\title{Fitting pseudo-\sersic\ (Spergel) light profiles to galaxies in interferometric data: the excellence of the $uv$-plane} 


   \author{Qing-Hua Tan
          \inst{1,2}
          \and
          Emanuele Daddi
          \inst{2}
          \and 
          Victor de Souza Magalh\~{a}es
          \inst{3,4}
          \and 
          Carlos G\'{o}mez-Guijarro
          \inst{2}
          \and
          J\'{e}r\^{o}me Pety
          \inst{3,5}
          \and
          Boris S. Kalita
          \inst{6,7,8}
          \and
          David Elbaz
          \inst{2}
          \and 
          Zhaoxuan Liu
          \inst{6,8,9}
          \and
          Benjamin Magnelli
          \inst{2}
          \and 
          Annagrazia Puglisi
          \inst{10,11}\thanks{Anniversary Fellow}
          \and
          Wiphu Rujopakarn
          \inst{12,13}
          \and
          John D. Silverman
          \inst{6,8,9}
          \and
          Francesco Valentino
          \inst{14,15}
          \and 
          Shao-Bo Zhang
          \inst{1}
          }

   \institute{Purple Mountain Observatory, Chinese Academy of Sciences, 10 Yuanhua Road, Nanjing 210023, People's Republic of China\\
            \email{qhtan@pmo.ac.cn}
        \and
            Universit\'{e} Paris-Saclay, Universit\'{e} Paris Cit\'{e}, CEA, CNRS, AIM, 91191, Gif-sur-Yvette, France
        \and
            Institut de Radioastronomie Millim\'{e}trique, 300 Rue de la Piscine, 38406 Saint-Martin d'H\`{e}res, France
        \and
        National Radio Astronomy Observatory, 800 Bradbury Dr., SE Ste 235, Albuquerque, NM 87106, USA    
        \and
        		LERMA, Observatoire de Paris, PSL Research University, CNRS, Sorbonne Universit\'es, 75014 Paris   
        \and
        		Kavli Institute for the Physics and Mathematics of the Universe, The University of Tokyo, Kashiwa, 277-8583, Japan
        	\and
        		Kavli Institute for Astronomy and Astrophysics, Peking University, Beijing 100871, People{\textquotesingle}s Republic of China
        	\and		
            Center for Data-Driven Discovery, Kavli IPMU (WPI), UTIAS, The University of Tokyo, Kashiwa, Chiba 277-8583, Japan
            \and 
            Department of Astronomy, School of Science, The University of Tokyo, 7-3-1 Hongo, Bunkyo, Tokyo 113-0033, Japan
        \and
            School of Physics and Astronomy, University of Southampton, Highfield SO17 1BJ, UK         
        \and
            Center for Extragalactic Astronomy, Department of Physics, Durham University, South Road, Durham DH1 3LE, UK
        \and
            National Astronomical Research Institute of Thailand, Don Kaeo, Mae Rim, Chiang Mai 50180, Thailand
        \and
            Department of Physics, Faculty of Science, Chulalongkorn University, 254 Phayathai Road, Pathumwan, Bangkok 10330, Thailand
        \and
            European Southern Observatory, Karl-Schwarzschild-Str. 2, D-85748 Garching bei Munchen, Germany
        \and
            Cosmic Dawn Center (DAWN), Denmark
             }


 
  \abstract
  {Modern (sub)millimeter interferometers, such as ALMA and NOEMA, offer high angular resolution and unprecedented sensitivity. This provides the possibility to characterize the morphology of the gas and dust in distant galaxies. To assess the capabilities of current softwares in recovering morphologies and surface brightness profiles in interferometric observations, we test the performance of the Spergel model for fitting in the $uv$-plane, which has been recently implemented in the IRAM software GILDAS (\texttt{uv$\_$fit}).
  Spergel profiles provide an alternative to the \sersic \ profile, with the advantage of having an analytical Fourier transform, making them ideal to model visibilities in the $uv$-plane. We provide an approximate conversion between Spergel index and \sersic \ index, which depends on the ratio of the galaxy size to the angular resolution of the data.
  We show through extensive simulations that Spergel modeling in the $uv$-plane is a more reliable method for parameter estimation than modeling in the image-plane,  as it returns parameters that are less affected by systematic biases and results in a higher effective signal-to-noise ratio (S/N). 
   The better performance in the $uv$-plane is likely driven by the difficulty of accounting for correlated signal in interferometric images. Even in the $uv$-plane, the integrated source flux needs to be at least 50 times larger than the noise per beam to enable a reasonably good measurement of a Spergel index. 
   We characterise the performance of Spergel model fitting in detail by showing that parameters biases are generally low (< 10\%) and that uncertainties returned by \texttt{uv$\_$fit} are reliable within a factor of two. Finally, we showcase the power of Spergel fitting  by re-examining two claims of extended halos around galaxies from the literature, showing that galaxies and halos can be successfully fitted simultaneously with a single Spergel model.
   }
  

  \titlerunning{Fitting pseudo-\sersic\ (Spergel) light profiles to galaxies in interferometric data}
  \authorrunning{Qing-Hua Tan et al.}

   \maketitle

%


\section{Introduction}\label{sec:intro}

Galaxy morphologies are closely linked to their formation and evolution \citep[e.g.,][]{conselice14}. One of the most effective ways to understand the structures of galaxies and how they evolve over time is by measuring the distribution of light within them, specifically in relation to their radial distance from the center \citep[e.g.,][]{tacchella15}.  To describe the morphological types of galaxies, two commonly used measurements are the half-light radius and the central concentration of light profiles (e.g., the \sersic \ index). These measurements provide insights into the shapes and sizes of galaxies and help classify them into different morphological categories.

Over the past decade, optical and near-infrared (IR) observations have allowed for the exploration of the structural properties of the stellar components in high-redshift star-forming galaxies \cite[e.g.,][]{wuyts11, vanderwel14,shibuya15,tacchella18,cutler22,chen22,kartaltepe23}. However, obtaining measurements of purely star-forming components based on ultraviolet (UV) and optical star formation rate (SFR) tracers are much more difficult, due to the effects of dust attenuation. At high redshifts, galaxy morphology can manifest in various ways, ranging from radial gradients \citep[][]{nelson16} to very complex asymmetrical distributions \citep{Le-Bail2023}. In massive ($M_* > 10^{10}\ M_\odot$) galaxies, most of the rest-frame UV/optical emission is re-emitted in the far-IR/submillimeter windows \citep[e.g.,][]{pannella15,wang19,fudamoto21,smail21,xiao23,gomez-guijarro23}, which provide an excellent alternative to obtain unbiased measurements of the morphology of star formation inside galaxies.

Recent developments in (sub)millimeter interferometers, such as ALMA and NOEMA, have made it possible to study the distribution of dust and molecular gas in high-redshift galaxies with high sensitivity \citep[e.g.,][]{barro16,hodge16,hodge19,tadaki17,elbaz18,fujimoto18,gullberg19,jimenez-andrade19,puglisi19,puglisi21,gomez-guijarro22a,stuber23}. This provides new observational constraints to the distribution of gas and dust-obscured star formation in galaxies. ALMA, with the widest array configurations, can potentially deliver very high spatial resolution of the order of a few tens of milliarcseconds (depending on frequency), which is unparalleled even by \textit{HST} and \textit{JWST} standards. ALMA has already revealed the complex structure of star formation on sub-kiloparsec scales in dusty star-forming galaxies \citep[e.g.,][]{iono16,gullberg18,hodge19,rujopakarn19}. Comparing star formation profiles with stellar mass profiles and morphology (more in general) is crucial for understanding the structural evolution of galaxies as a result of star formation \citep[e.g.,][]{cibinel15}. To extract the full amount of information contained in the data, it is also essential to have appropriate technical tools. For example, ALMA and NOEMA interferometers gather data in the form of {\it visibilities} between antennas in the $uv$-plane, unlike the optical data that are directly images from CCDs or other electronic detectors.

So far, modeling galaxy morphology profiles in the submillimeter, based on interferometric data, has typically involved reconstructing images from the {\it visibilities}, followed by featuring \sersic \ (or other kind of) profiles fitting.  
This approach has been necessary due to the lack of an effective way to fit generic \sersic \ profiles in the $uv$-plane, as  supported by common software packages dedicated to the calibration and analysis of radio-interferometric data  \citep[e.g., CASA, AIPS, MIRIAD, GILDAS; a summary of profiles provided by each software package in the visibility modeling is given by][]{marti-vidal14}. 

Visibilities are the immediate output of interferometeric observations that sample the Fourier transform of the sky brightness distribution and consist of amplitudes and phases (depending on angles between antenna pairs as projected on the sky). The amplitudes and phases can be represented as imaginary numbers in the $uv$-plane, with their $uv$ elongation depending on antenna separation and frequency. As the Earth rotates, each pair of antennas in the interferometric array will trace out an elliptical track in the $uv$-plane. The Fourier transform of the measured visibilities produce a dirty image. The  point-spread function (PSF) of the resulting image, known as {\it dirty beam}, has a complex shape with, even in case of relatively good sampling of the $uv$-plane, numerous positive and negative sidelobes, extending to large spatial scales. 

As each data element in the $uv$-plane (visibility) affects the data at all spatial scales via the Fourier transform, the signal as well as the noise in the resulting images are always strongly correlated, especially on scales according to the full width at half maximum (FWHM) of the dirty beam. This correlation can significantly impact the measurements of sources' structure \citep[e.g.,][]{condon97,marti-vidal14,pavesi18,tsukui23}. When analyzing image-based measurements, it has been shown that ignoring signal correlation in interferometric images can lead to a significant underestimation of statistical uncertainties and, consequently,  misinterpreted results \citep{tsukui23}. However, many recent studies \citep[e.g.,][]{elbaz18,fujimoto18,hodge19,lang19,fudamoto22} measured submillimeter morphologies in the image plane using tools such as {\it galfit} \citep[][]{peng02,peng10}. These tools are optimised for optical/near-IR observations and, as such, do not account for the complex noise correlation typical of interferometric observations.

Interferometric images are often cleaned during deconvolution, which replaces the dirty beam with well-defined PSFs. However, these routines are based on strong assumptions and are not fully objective. Moreover, they do not account for the strong correlation between pixels on the scale of the beam. For example, high-fidelity observations of $z\sim 3$ star-forming galaxies by \citet{rujopakarn19} using ALMA  show that the small-scale source structures of galaxies are affected by both the weighting scheme and the cleaning (deconvolution) algorithm in imaging procedures.

In contrast, working directly in the $uv$-plane would be preferable because the measured data points are independent there-in. For example, it has been suggested that analyzing observations of very compact sources in Fourier space is more reliable than image-based analyses \citep{marti-vidal12}. However, the lack of tools that allow general profile fitting in the $uv$-plane is an obstacle. Typical codes allows for Gaussian fitting, in addition to the standard PSF fitting, and sometimes exponential profiles (i.e., the particular case of $n=1$ for the \sersic \ profile).

One possible way to approximate the standard surface brightness profile (i.e., general  \sersic) of galaxies is to use linear superpositions of Gaussians \citep{hogg13}. However, this approach has limitations. For example, it is difficult to precisely measure how the light is concentrated in the center of  galaxies, and the comparison to results obtained in optical/near-IR bands becomes prohibitive. This is because general \sersic \ profiles are not analytically transformable into Fourier space, making it computationally intensive to fit a \sersic \  profile to visibilities that requires large numbers of numerical Fourier transforms while fitting the model to the data.

Quite conveniently, the issue with the lack of analytical Fourier-transformability of the \sersic \  profile has already been addressed in the framework of optical imaging, where it became   necessary to account for spatially and temporally varying PSFs, that requires computationally intensive convolutions. \citet{spergel10} proposed a solution based on the incomplete Bessel function of the third kind. This function closely approximates \sersic \ functions and is commonly known as the Spergel profile (see Section~\ref{subsec:spergel} for a description of the profile).
The Spergel profile has been recently also incorporated into the MAPPING procedure of GILDAS \footnote{\url{http://iram.fr/IRAMFR/GILDAS}} \citep{guilloteau00}. This allows for the study of galaxies morphology in interferometric observations with unprecedented detail, enabling comparison to optical studies. The idea is to model  galaxy structures accurately using functions that approximate the \sersic \ profile in the $uv$-plane (i.e., pseudo-\sersic). 

Modeling the galaxy submillimeter structure using a Spergel profile has been discussed for selected examples of high-redshift star-forming galaxies \citep[][]{kalita22,rujopakarn23}.
However, despite the novelty of the exercise, there have been no attempts in the literature to systematically characterise the performances of Spergel modeling of light profiles of galaxies in the $uv$-plane. This includes the returned fidelity and accuracy in parameter estimation, required signal-to-noise ratios (S/N) for attempting complex modeling, and possible degeneracies between parameters. This is similar to what has been extensively done in the optical images for {\it galfit} or other tools during the last decades \citep[e.g.,][and many others]{moriondo00,pignatelli06,haussler07,mancini10,hoyos11,hiemer14,lange16,tortorelli23}. Additionally, the conversion of Spergel indices to Sersic indices, which enables the comparison of submillimeter/millimeter measurements of Spergel profiles to those derived from the classic approach of galaxy light profile modeling \citep[i.e., {\it galfit};][]{peng02}, was only briefly explored by \citet{spergel10}. 

In this paper, we aim to provide a technical assessment of the use of the Spergel profile. Our investigation focuses on the robustness of profile fitting with Spergel models using simulated data with real noise from actual observations, and compares it with the traditional application of profile fitting on interferometric reconstructed images. We plan to use these results to perform Spergel modeling on a large galaxy sample taken from the ALMA archive in a forthcoming paper (Q. Tan et al. in preparation).

\begin{figure*}[tbp]
\centering
\includegraphics[width=0.49\linewidth]{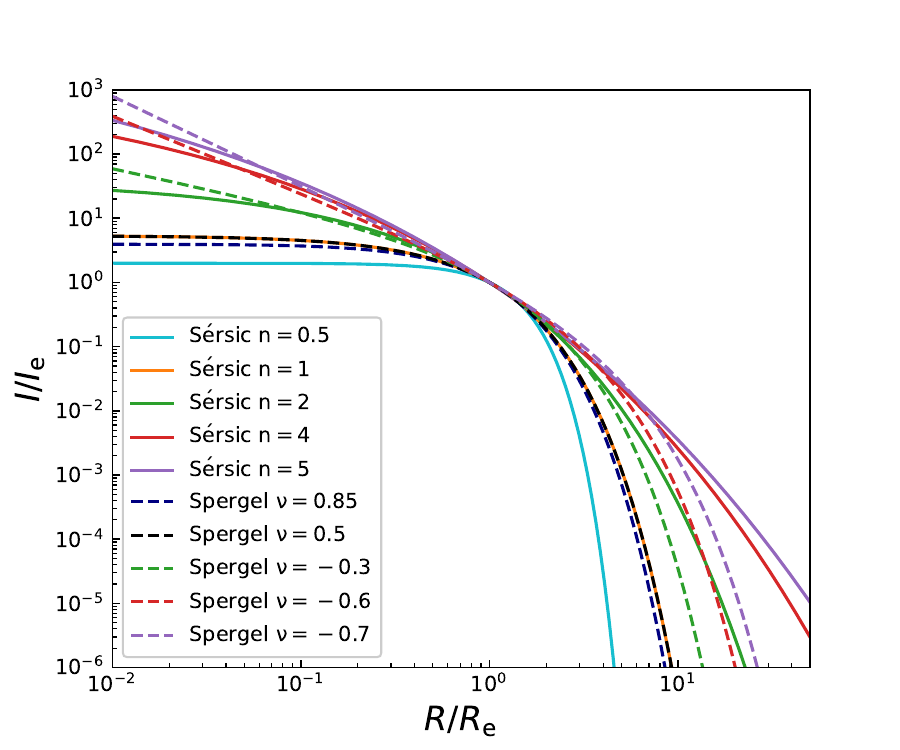}
\includegraphics[width=0.49\linewidth]{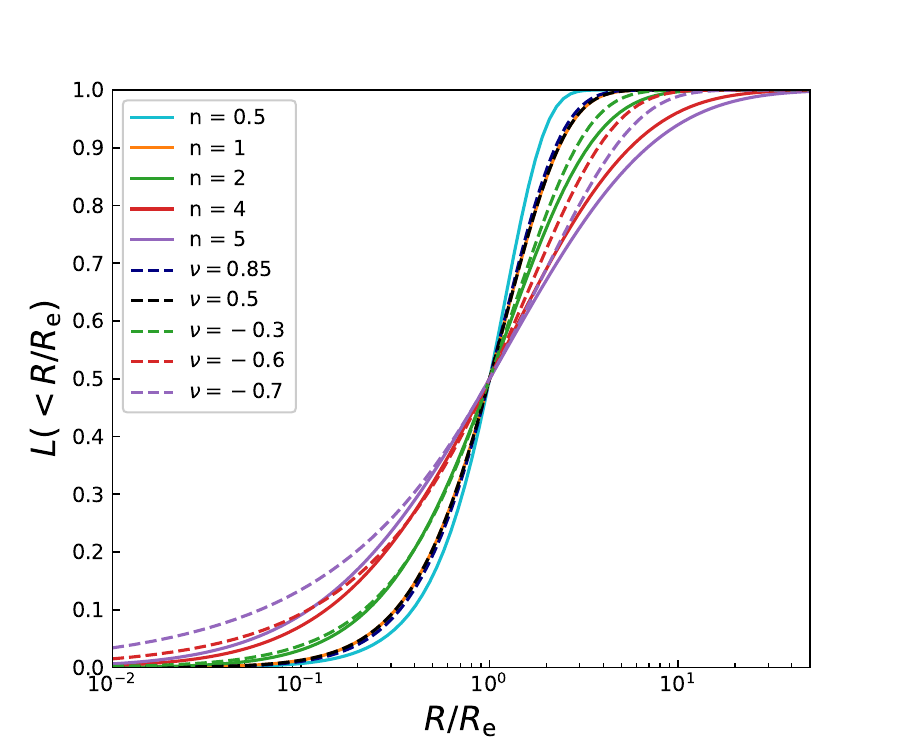}
\caption{{\bf Left}: Comparison of surface density profiles for \sersic \ (solid lines) and Spergel (dashed lines) functions. The light profile with \sersic \ index of 0.5, 1, 2, 4, and 5, and Spergel index of 0.85, 0.5, -0.3, -0.6, and -0.7, are shown in different colors. {\bf Right}: Comparison of the cumulative distributions for \sersic \ (solid lines) and Spergel (dashed lines) functions. By definition, $R_{\rm e}$, contains half of the integrated galaxy light. }\label{fig:profiles}
\end{figure*}

The paper is structured as follows. In Section~\ref{sec:functions}, we recall the definition of the radial profile functions used to constrain the light profile of galaxies in the image-plane (\sersic) and $uv$-plane (Spergel), respectively.
We compare the two profiles and discuss how their parameters can be converted from one to the other, for this comparison. Section~\ref{sec:simulation} introduces the method used to generate the simulated data, with a description of the \texttt{uv$\_$fit} algorithm. In Section~\ref{sec:comparison}, we present the results of a study that tested the robustness of profile fitting in the $uv$-plane against model fits to image data. This includes a comparison of the recovery of the structural parameters and the accuracy of the measurements. The analysis of the absolute accuracy of $uv$-plane modeling, the reliability of parameter uncertainties, and the covariance of fitted parameters are described in Section~\ref{sec:spergel_abs}. In Section~\ref{sec:discussion}, we discuss the simulation results and explore possible reasons for the different measurement performances of interferometric data. We also discuss the implication for the study of galaxy morphology,  using an example based on re-examining previously published ALMA data. The conclusions of the paper are presented in Section~\ref{sec:conclusion}.


\section{\sersic\ and Spergel radial profile functions} \label{sec:functions}

In this section, we first recall the definitions of the \sersic\ and Spergel profiles. Then we proceed to a qualitative comparison between them as their indices vary, emphasizing the role played by the  spatial scales actually observed.  Finally, we present  empirical recipes to convert Spergel indices to their equivalent \sersic\ ones, as well as effective radii and total fluxes. These recipes will be used throughout the paper to directly compare Spergel and \sersic\ fits under realistic noise conditions from our simulations.

\subsection{The S${\rm \acute{e}}$rsic profile} \label{subsec:sersic}

As a generalization of the $r^{1/4}$ law, the $r^{1/n}$ profile first proposed by \citet{sersic68} is one of the most common functions used to describe how the intensity of a galaxy varies with distance from its center.
The surface density (or equivalently surface brightness) of the \sersic\ profile can be written as
\begin{equation}
\Sigma(R) = \Sigma_{\rm e}{\rm exp}\left[-\kappa \left( \left( \frac{R}{R_{\rm e}} \right)^{1/n} - 1 \right)\right ],
\end{equation}
where $R$ is the projected distance to the source center, $R_{\rm e}$ is the effective radius containing half of the total luminosity, $\Sigma_{\rm e}$ is the surface brightness at $R_{\rm e}$, and $n$ is the \sersic \ index which determines the shape of the light profile (see Fig.~\ref{fig:profiles}). The parameter $\kappa$ is a function of \sersic \ index and is such that $\Gamma (2n) = 2\gamma (2n, \kappa)$, where $\Gamma$ and $\gamma$ represent the complete and incomplete gamma functions \citep{ciotti99}, respectively. We will use the terms half-light radius or effective radius interchangeably to refer to the radius within which half of a galaxy's luminosity is contained.

The \sersic \ index, $n$, determines the degree of curvature of the profile, with $n=0.5$ giving a Gaussian profile, $n=1$ an exponential disk profile and $n=4$ generally associated with galaxy bulges. As the index $n$ increases, the core steepens more rapidly for $R<R_{\rm e}$, and the intensity of the outer wing at $R>R_{\rm e}$ is significantly extended. 

However, as mentioned in the introduction, the general \sersic \ profile is not analytically transformable in Fourier space for most values of the parameter $n$, as the parameter $\kappa$ cannot be solved in closed form. Various techniques have been developed to address this issue when calculations of the Fourier transform are needed, including numerical integration methods, approximations, and asymptotic expressions \citep[e.g.,][]{ciotti99,mazure02,baes11}. The inability to solve the \sersic \ profile analytically in Fourier space creates challenges in certain scenarios. For example, when performing convolutions, such as corrections for seeing, or when directly working in Fourier space, like with interferometric data.

\subsection{The Spergel profile} \label{subsec:spergel}

\citet{spergel10} introduced an alternative to the \sersic \ model for galactic luminosity profiles with functional form
\begin{equation}
\Sigma_\nu (R) = \frac{c_\nu^2 L_{\rm 0}}{R_{\rm e}^2} f_\nu \left( \frac{c_\nu R}{R_{\rm e}} \right)\label{eq:spergel}
\end{equation}
where $f_\nu(x) = \left( \frac{x}{2} \right)^\nu \frac{K_\nu (x)}{\Gamma (\nu + 1)}$, $\Gamma$ is the Gamma function, $K_\nu$ is a modified spherical Bessel function of the third kind, $c_\nu$ is a constant, $R_{\rm e}$ is the half-light radius, and $\nu$ is known as Spergel index that controls the relative peakiness of the  core and the relative prominence of the  wings (similar to \sersic \  $n$), with a theoretical limit of $\nu > -1$.

This family of functions is found to provide a good fit for galaxy light profiles and resembles the \sersic \ function over a range of indices. The Spergel profile at $\nu=0.5$ is identical to an exponential profile, which is equivalent to a \sersic \ profile with $n=1$. However, the two functions do not exactly coincide for different Spergel $\nu$ (see Fig.~\ref{fig:profiles}). 

The Spergel profile has a significant advantage over the \sersic \ profile because it is analytic in both real space and Fourier space \citep[][]{spergel10}, which means that it can be described mathematically using equations, making it easier to work with and analyze in the $uv$-plane.

\subsection{Qualitatively relating the \sersic \ and Spergel profiles}\label{subsec:profile-relation}

In the left panel of Fig.~\ref{fig:profiles}, a comparison is shown between \sersic \ and Spergel profiles. This comparison covers a range of $n$ and $\nu$ indices. All the profiles are normalized at $R_{\rm e}$. 
Within a certain range of $\nu$, the Spergel profiles resemble \sersic \ profiles in shape such that a relation can be conceived between $\nu$ and $n$. For example, the Spergel profile at $\nu=-0.6$ and in the radial range near the effective radius, exhibits similarities to a {\it de\ Vaucouleurs} $n=4$ profile \citep[see also][]{spergel10}. However, when elongating far from the normalization point, these profiles start to differ at both the innermost (i.e., $R < 0.1\ R_{\rm e}$) and outermost (i.e., $R > 5\ R_{\rm e}$) regions, indicating that converting one index into the other depends on the observed scales. This is generally the case for all  \sersic \ models with $n>1$, respect to their first-order matching Spergel profiles with $-1<\nu<0.5$: they display steeper inner profiles and drop faster at large radii. This is discussed more further down in this Section.

We also see from Fig.~\ref{fig:profiles} that Spergel models are not meant to reproduce \sersic\ models with a flatter shape than $n=1$, i.e., $n<1$. For example, even choosing $\nu=0.85$ the Spergel model only slightly flattens compared to the $\nu=0.5$ ($n=1$) case, and it remains far from resembling a Gaussian model $n=0.5$. Given that fitting Gaussian models in the $uv$-plane is a straightforward approach, we suggest using this method instead of attempting Spergel fits with  high values of $\nu>1$.

In the right panel of Fig.~\ref{fig:profiles}, we show  cumulative distributions for the \sersic \ and Spergel profiles. When truncated at $10\ R_{\rm e}$, an exponential profile (\sersic \ $n=1$ and Spergel $\nu=0.5$) retains almost all of its flux ($\sim 100\%$). In contrast, a \sersic \ profile with $n=4$ retains 96.1$\%$ of its flux, while 99.7$\%$ of the flux is contained within this radius for a Spergel profile with $\nu= -0.6$. At the small radius, within the inner $0.05\ R_{\rm e}$, an exponential profile contains 0.3$\%$ of the flux, while $n=4$ contains 3.2$\%$ of the flux. For a Spergel profile with $\nu= -0.6$, 5.4$\%$ of the flux is contained within this radius. As the index $n$ increases (or decreases for $\nu$), the differences between the fluxes contained within the radius of the small and large ends also somewhat increase, while remaining overall contained. A comparison of the \sersic \ profile with the Spergel profile using mathematical simulations is presented in Appendix~\ref{app:analytic-matching} (see Fig.~\ref{fig:app-profile-mcmc}).

\begin{figure*}[tbp]
\centering
\includegraphics[width=\linewidth]{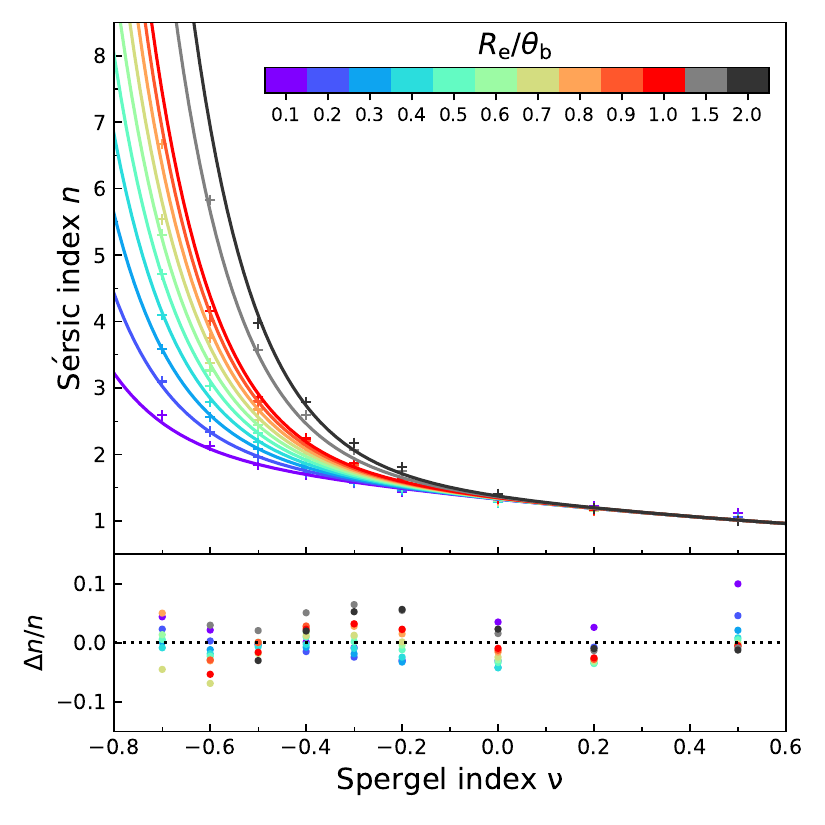}
\caption{Comparison  S${\rm \acute{e}}$rsic and Spergel indexes. 
The data (crosses) in the top-panel represent the {\it galfit} S${\rm \acute{e}}$rsic indices measured for sources created with input Spergel model, for a range of $\nu$ values from $-0.7$ to 0.5, and for different source effective radii (expressed in terms of the FWHM of the synthesized beam, $R_{\rm e}/\theta_{\rm b}$) ranging from 0.1 to 2.0. The solid curves in the top panel denote the best-fit empirical relation between Spergel $\nu$ and {\it galfit} \sersic \ $n$, depending on $R_{\rm e}/\theta_{\rm b}$, as expressed by Eq.~(\ref{eq:conversion}). The lower panel displays normalised residuals to the proposed relation.}\label{fig:index-matching}
\end{figure*}

\begin{figure*}[tbp]
\centering
\includegraphics[width=0.45\linewidth]{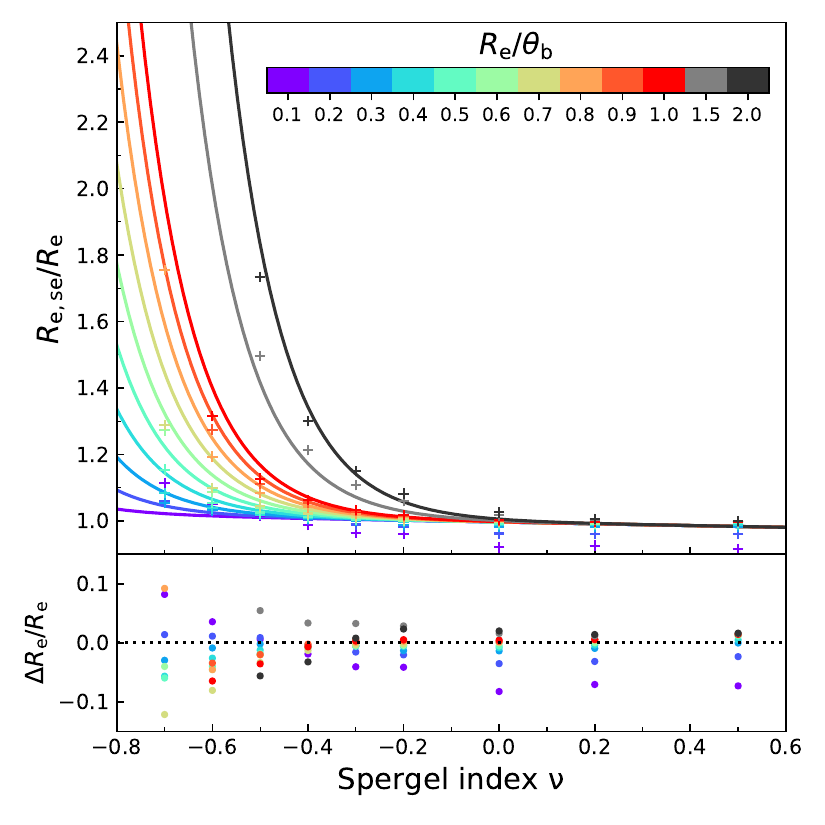}
\includegraphics[width=0.45\linewidth]{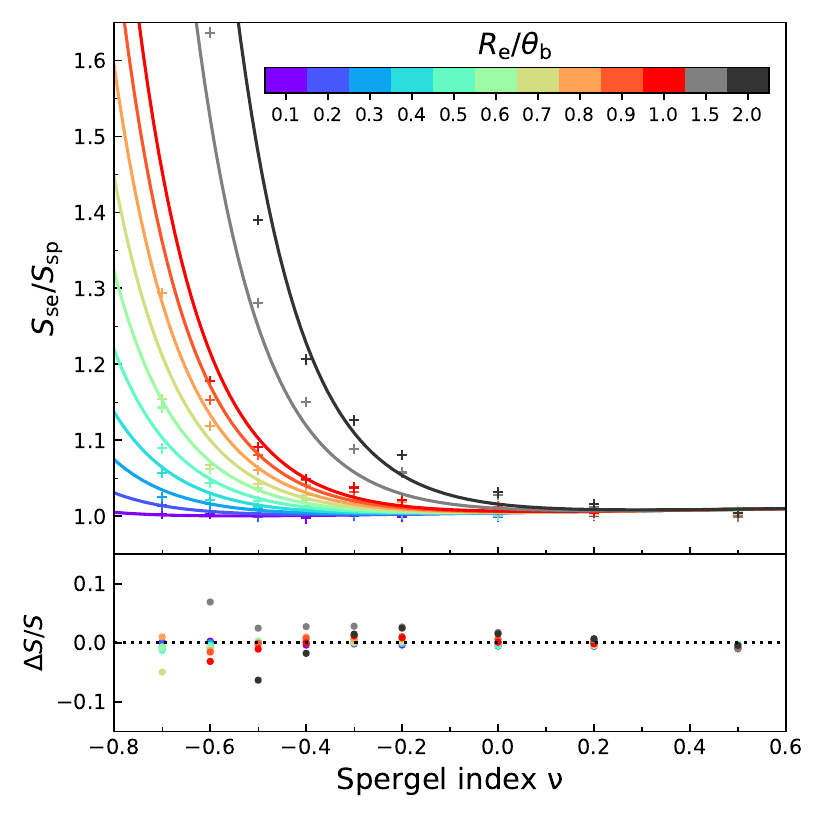}
\caption{Similar to Fig.~\ref{fig:index-matching}, but showing the ratio of effective radii ({\it left}) and total fluxes ({\it right}) obtained from fitting S${\rm \acute{e}}$rsic models simulated galaxies created using Spergel profiles (crosses), plotted as a function of their Spergel index. The solid curves denote the best-fit empirical relations as expressed by Eq.(\ref{eq:ratio-size-flux}). The bottom panels show normalised residuals to the best-fit, which are largely within 10\%. }
\label{fig:infinite-simu}
\end{figure*}

\subsection{Converting the Spergel index into the equivalent \sersic \ index}
\label{subsec:index-conversion}

To empirically derive the conversion (spatial-scale dependent) between  Spergel $\nu$ and the \sersic \ $n$, we create noise-free\footnote{{ The model sources were created with an extremely high S/N  to ensure a robust {\it galfit} measurement}} images of Spergel two-dimensional models, as described in the next section, and use {\it galfit} to measure the corresponding \sersic\ index. 
In Fig.~\ref{fig:index-matching}, the cross symbols represent the {\it galfit} measurements, which we refer to the intrinsic best value and are labelled as $n_{\rm galfit}$, compared to the input $\nu$ for different source sizes of $R_{\rm e}/\theta_{\rm b}$ ranging from 0.1 to 2.0. Here $\theta_{\rm b}$ is defined as the synthesized circularized beam size, given by $\sqrt{ab}$, where $a$ and $b$ represent the FWHMs of the major and minor axes of the synthesized beam, respectively. 

As a zero-order check of the procedure, Fig.~\ref{fig:index-matching} shows that for a Spergel profile with $\nu=0.5$, sources with different sizes all return a \sersic \ index of $n=1$ in {\it galfit}, as expected. This result is accurate within $3\%$, which represents the inherent maximal precision of this empirical calibration. We find that $\nu=-0.6$ corresponds to $n=4$ for R$_{\rm e}/\theta_{\rm b}\sim0.9$--1.0, which is close to when the FWHM scale of the galaxy and of the beam are identical, as expected \citep{spergel10}.
However, when R$_{\rm e}/\theta_{\rm b}<0.5$ an input $\nu= -0.6$ converts to $n\sim2$--3, while models with $\nu= -0.6$ and R$_{\rm e}/\theta_{\rm b}>1$ correspond to steeper $n>4$. The dependence of the conversion  on R$_{\rm e}/\theta_{\rm b}$ systematically decrease with increasing the Spergel indices, and nearly vanishes for $\nu>0$ (Fig.~\ref{fig:index-matching}).

We find that the whole set of measurements can be well described by the form:
\begin{equation}
n(\frac{R_{\rm e}}{\theta_{\rm b}}, \nu) \sim p_1 \frac{R_{\rm e}}{\theta_{\rm b}} {\rm exp}(p_2 \nu) + p_3 \nu^2 + p_4 \nu + p_5
\label{eq:conversion}
\end{equation}
for  sources with $R_{\rm e}/\theta_{\rm b}$ ranging from 0.1 to 2.0, respectively. The solid lines in Fig.~\ref{fig:index-matching} represent the best-fit relation with coefficients $p_1 = 0.0249$, $p_2 = -7.72$, $p_3 = 0.191$, $p_4 = -0.721$, and $p_5 = 1.32$. The residuals of the fit indicate that the uncertainties of \sersic \ $n$ are largely within 10\% for model sources with $R_{\rm e}/\theta_{
\rm b}$ of 0.1--2.0 at $\nu>-0.7$ (see Fig.~\ref{fig:index-matching}). We confirm this trend  when comparing the Spergel index with the \sersic \ index analytically through a mathematical matching between the \sersic \ and Spergel functions (see Fig.~\ref{fig:app-match-mcmc}).

This analysis demonstrates that converting a Spergel index to a \sersic\ index depends on the ratio between the angular size of the galaxy and that of the beam of the observations being examined. We have mapped this relationship over a reasonably large range of this ratio, i.e., $R_{\rm e}/\theta_{
\rm b}$ of 0.1--2.0. Anticipating results from forthcoming paper, which focuses on analyzing the morphologies of an ALMA archival sample of about 100 distant star-forming galaxies in the submillimeter bands (Q.~Tan et al., in preparation), we find that around 82\% (93\%) sources fall within the range of $R_{\rm e}/\theta_{\rm b} = 0.1-1.0 \  (0.1-2.0)$, while approximately 88\% (99\%) of sources exhibit best-fitting $\nu > -0.7$ ($\nu < 1$). The median $R_{\rm e}/\theta_{\rm b}$ is 0.32  and the semi-interquartal range is 0.2--0.6.
This suggests that the calibration presented in Fig.~\ref{fig:index-matching} is representative of general   observations of distant galaxies with ALMA and NOEMA.

\subsection{Converting half-light radii and total fluxes from Spergel to \sersic}

The differences in the profiles imply that also for sizes (half-light radii) and total fluxes, a conversion might be required when comparing results based on Spergel to those from \sersic. 
Based on our simulations, we find that both the size and flux density estimated by {\it galfit} using a \sersic \ profile tend to be larger than when using Spergel, as the \sersic \ index  becomes larger. Instead,  axis ratios and position angles are unaffected. Fig.~\ref{fig:infinite-simu} shows the ratios of $R_{\rm e}$ and flux densities measured from  fitting  \sersic \ models to simulated galaxies created as Spergel models, as a function of Spergel index. Both the ratio of $R_{\rm e}$ and flux density exhibit a similar increasing trend as the profile get steeper. The ratio  becomes larger for better resolved sources  (i.e., larger $R_{\rm e}/\theta_{\rm b}$). For example, in the case of a large-sized ($R_{\rm e}/\theta_{\rm b}=1$) galaxy with a {\it de Vacouleurs}-like profile of $\nu=-0.6$, the $R_{\rm e}$ estimated by the \sersic \ model can be larger by about 30--40\%, while the excess flux density is around 15--20\%. By comparison, the relative increase in $R_{\rm e}$ and flux density are less significant for sources with flatter profile or smaller sizes compared to the beam. 

To correct for these systematical biases and thus to enable comparison of measurements using Spergel and \sersic \ profiles, we fit the distribution of the ratio of half-light radius and total flux measured from \sersic \ to the Spergel-based input values. To distinguish the fitted parameters between those obtained from Spergel versus \sersic \ profile modeling, we label $R_{\rm e}$ and $S_{\rm sp}$ the half-light radius and total flux measured from Spergel profile fitting, respectively. For \sersic-based measurements, we use instead $R_{\rm e,se}$ and $S_{\rm se}$. We find that both ratios of $R_{\rm e,se}/R_{\rm e}$ and $S_{\rm se}/S_{\rm sp}$ can be accurately described by a similar form: 
\begin{equation}
    r(\frac{R_{\rm e}}{\theta_{\rm b}}, \nu) \sim p_1 (\frac{R_{\rm e}}{\theta_{\rm b}})^2 {\rm exp}(p_2 \nu + p_3 \frac{R_{\rm e}}{\theta_{\rm b}}) + p_4 \nu + p_5
    \label{eq:ratio-size-flux}
\end{equation}
for  sources with $R_{\rm e}/\theta_{\rm b}$ ranging from 0.1 to 2.0. For the size ratio of $R_{\rm e,se}/R_{\rm e}$, the best-fit gives coefficients $p_1 = 0.00138$, $p_2 = -8.96$, $p_3 = 0.260$, $p_4 = -0.0260$, and $p_5 = 0.996$, while for the flux ratio of $S_{\rm se}/S_{\rm sp}$, the best-fit coefficients are $p_1 = 0.00217$, $p_2 = -7.43$, $p_3 = 0.149$, $p_4 = 0.00942$, and $p_5 = 1.00$ (see the solid lines in Fig.~\ref{fig:infinite-simu}). 
Analytical  calculations, matching the Spergel profile with \sersic \ profiles numerically (see Appendix~\ref{app:analytic-matching}),  fully confirm the trends encoded in the Eq.(\ref{eq:ratio-size-flux}) above and in  Fig.~\ref{fig:infinite-simu}.
\bigskip

We caution that while we believe that our methodology captures the bulk of the systematic effects in the conversion as encoded in the $R_{\rm e}/\theta_{\rm b}$ ratio,  some further systematics might be expected depending on higher order terms describing the actual shape of the beam. We derived best fitting parameters for Eq.(\ref{eq:conversion}) and Eq.(\ref{eq:ratio-size-flux}) averaging over three different ALMA array configurations. By comparing to results from single ALMA array configurations, we estimate that further systematic uncertainties are small. The details of the three ALMA array configuration are summarized in Table~\ref{tab:configurations} (see Appendix~\ref{app:configurations}).

\section{Technical aspects: simulating galaxies under realistic noise conditions, and measuring their properties}\label{sec:simulation}

\subsection{Realistic noise map}\label{subsec:noise-map}

Each simulated galaxy was created by inserting a  model source signal into an empty dataset with realistic noise. The noise was obtained from real data using ALMA band 7 observed visibilities from galaxies in a recent survey \citep[e.g.,][]{puglisi19,puglisi21,valentino20}. 

To analyze the data, the calibrated ALMA visibilities for several targets were exported from CASA using the \texttt{exportuvfits}. The exported data was then converted to  $uv$-tables using the GILDAS \texttt{fits$\_$to$\_$uvt} task. The four spectral windows were combined using the \texttt{uv$\_$continuum} and \texttt{uv$\_$merge} tasks. Prior to introducing the model source into the $uv$-plane, any detected sources were subtracted from the visibility data using the best-fitting Spergel model to the visibilities. This produced a residual data set which, upon inspection, did not reveal any further source. 

\begin{figure*}[tbp]
\centering
\includegraphics[width=0.48\linewidth]{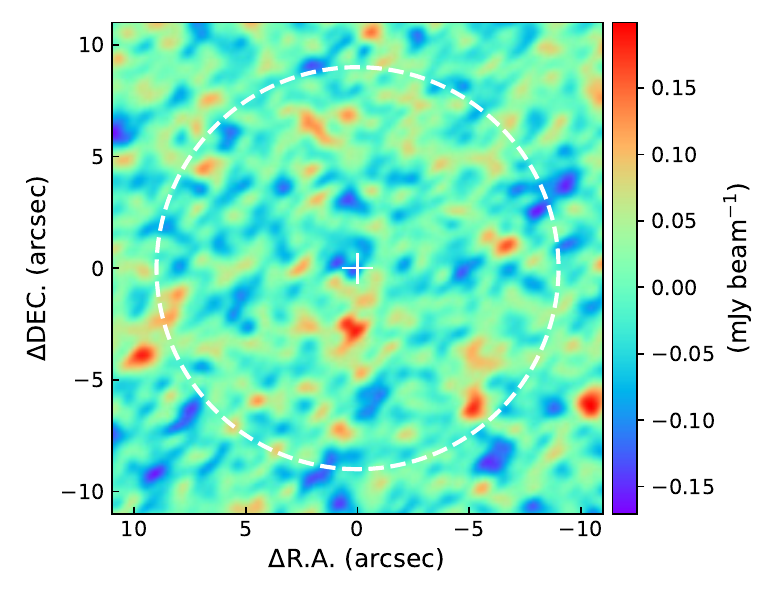}
\includegraphics[width=0.4\linewidth]{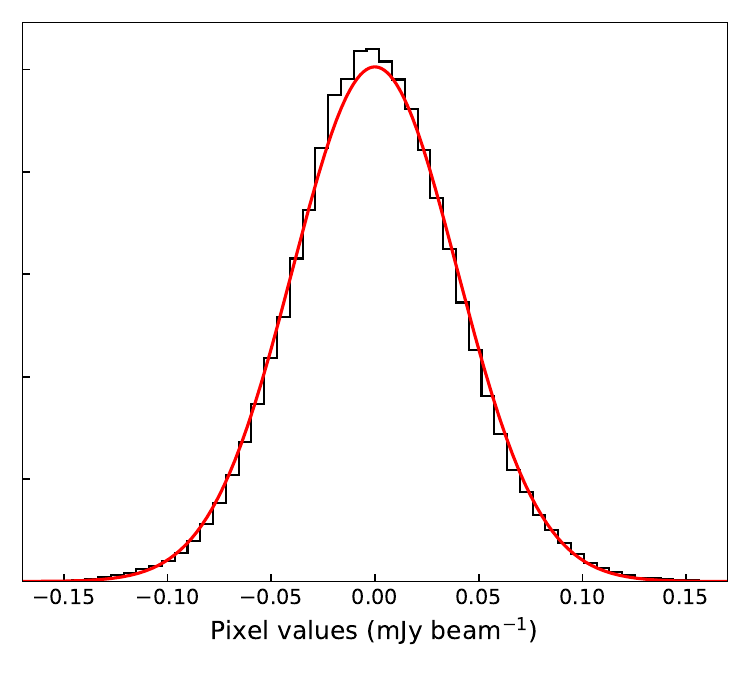}
\caption{{\bf Left}: One of the empty datasets used for our simulations, after subtracting a source (which was close to the phase center)  from the visibilities. The dashed white circle represents the primary beam, i.e., the FOV of the data set. {\bf Right}: Histogram  of  pixel values within the primary beam. The red line represents the best-fitting  Gaussian model. The fit suggests that the noise data is at least approximately Gaussian.}\label{fig:residual}
\end{figure*}

Fig.~\ref{fig:residual} (left) shows the map of the residual $uv$ data for a case in our simulations: the primary beam field of view (FOV) is 18$''$, and the FWHM of the synthesized beam using natural weighting is $\theta_{\rm b}=1.0''$. The right panel of Fig.~\ref{fig:residual} shows the pixel distribution, which can be fitted well with a Gaussian profile. This indicates that the data is mostly noise without any other significant features. The noise level measured from the Gaussian fit is 40 $\mu$Jy beam$^{-1}$. We refer to the rms noise derived in this way as $\sigma_{\rm b}$ in the following sections. This is an objective characterization of the noise in the data, independent on source properties. 

We emphasize that each empty galaxy map can be used for a large number of independent simulations. This can be achieved by placing simulated signal at random positions within the primary beam. Considering that half-synthesized-beam offsets produce an independent noise realization, the number of such independent realizations are of order 1000 for each empty ALMA band-7 dataset. 

\subsection{Source models}\label{subsec:source-model}

The model sources were created using GILDAS through the MAPPING task \texttt{uv$\_$fit}. We generate elliptical Spergel model sources by fixing their  seven free parameters: centroid position, flux density, effective radius $R_{\rm e}$, minor-to-major axis ratio, position angle (PA), and Spergel index $\nu$\footnote{In practice this is done by fitting to the empty data a model with all parameters fixed and negative flux. The residual image will then have the desired model added (and positive).}. The flux density varies over a range with step sizes of factors of two. This range corresponds to a total flux density that is normalized by the noise, varying from 25 to 400, which is represented by the ratio of integrated flux density to the pixel rms noise (S$_{\rm tot}$/$\sigma_{\rm b}$). We define S/N as the ratio of the integrated flux density to the noise per beam S$_{\rm tot}$/$\sigma_{\rm b}$: the advantage of this choice lays in its model-independence and reliance on very basic properties of the source and of the noise. Note that for extended sources, the S/N defined in this way is obviously higher than the S/N eventually recovered for the integrated flux density coming from a full Spergel/\sersic \ profile. 

The simulation process involves setting the size of the sources in units of $\theta_{\rm b}$, i.e., $R_{\rm e}/\theta_{\rm b}$ = 0.1, 0.2, 0.4, and 0.7, to represent very compact, small, intermediate, and large-sized (relative to the beam) sources, respectively. For each set of Monte Carlo (MC) sources that share a fixed effective radius and axis ratio $q$ ($\equiv b/a$), the Spergel model sources vary in both their flux density (thus, S/N) and Spergel index with values of $\nu=$ 0.5, $-0.3$, $-0.5$, $-0.6$, and $-0.7$. The position angle remain constant. To mimic real observations, we add the model sources to the realistic noise data that is derived from observed visibilities, to produce a simulated data set (see Fig.~\ref{fig:residual} for an example).

\subsection{Visibility model-fitting with Spergel profile}

Fitting   elliptical Spergel models to the simulated dataset is performed using the task \texttt{uv$\_$fit}. All the seven fitted parameters are allowed to vary. As typically done within MAPPING's \texttt{uv$\_$fit} in GILDAS, we generate a  range of initial guesses for each fitted parameter, which are built into a N-dimensional list of combinations of these guesses. This approach helps to explore a wider range of possible solutions and thus identify the best fit model and return a well-sampled range for uncertainties. We test the fit by setting the starting range parameters of initial guesses within a factor of 2 centered on the input mock values and the number of start parameters (e.g., 3 guesses for each parameter). We found that for simulated sources (where  model parameters are known a priori), the results using single initial guesses identical to the real parameters are not significantly different from when the code is run using multiple initial guesses for each parameter. However, using \texttt{uv$\_$fit} with a large range of initial guesses is critical for real observations, where the source's properties are not known beforehand.

\begin{figure*}[tbp]
\centering
\includegraphics[width=0.75\linewidth]{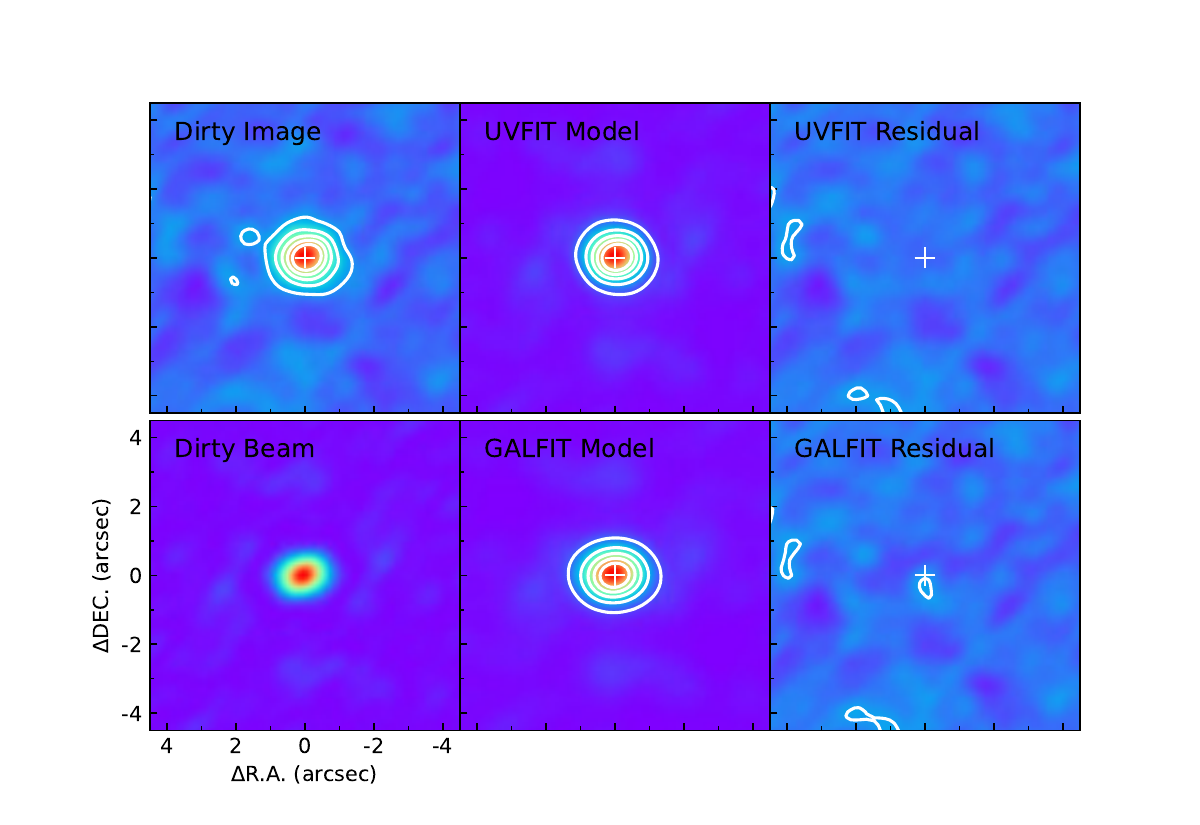}
\caption{Example of a simulated source generated with a Spergel profile of $S_{\rm tot}/\sigma_{\rm b}=$50, $R_{\rm e}/\theta_{\rm b}$=0.4, axis ratio of $q$ = 0.75, position angle of PA = 30$^\circ$, and Spergel index of $\nu = -0.6$. From left to right: the dirty image (top-left), dirty beam (PSF used for convolution in {\it galfit}; bottom-left), best-fit source models convolved with the dirty beam (middle), and residuals after subtracting the model source (right). The model source and the model-subtracted residual shown in the top and bottom rows were derived from the \texttt{uv$\_$fit} and {\it galfit} fits, respectively. Each image cutout is $9''\times 9''$. The contours start from 2$\sigma$ and increase in steps of 4$\sigma$. White crosses mark the best-fit source position obtained from \texttt{uv$\_$fit}.}\label{fig:model-residual}
\end{figure*}

\subsection{Image model-fitting with \sersic \ profile}

Each  $uv$-plane simulation is imaged  and then fitted with a single-component \sersic \ profile using {\it galfit} \citep{peng02}. 
The {\it galfit} run is performed on the dirty maps, which are created by Fourier transforming visibilities without cleaning. The (full) dirty beam is  used as the {\it galfit} PSF. The known parameters of the mock sources are used as initial guesses for {\it galfit}. All  parameters are left free without constraints in the fitting. The initial guess of the \sersic \ $n$ is calculated by converting the input $\nu$ using Eq.~(\ref{eq:conversion}). The \sersic \ fits also provide measurements of seven free parameters: central position, total magnitude, effective radius, \sersic \ index, axis ratio, and PA. In some cases, {\it galfit} may fail to provide accurate measurements. To ensure the validity of our comparisons, we remove measurements in both the image- and $uv-$plane for sources with output parameters marked as problematic in {\it galfit}. All this procedure is extremely favorably biased towards positively amplifying the performances of  {\it galfit}.
 
 Fig.~\ref{fig:model-residual} illustrates a typical case of a simulated source generated with a Spergel profile and with $S_{\rm tot}/\sigma_{\rm b}$ ratio of 50. Additionally, it shows the best-fit models derived from \texttt{uv$\_$fit} and {\it galfit}, respectively. Both methods provide good constraints to the simulated source at this S/N ratio, as no significant component is visible in the residual map.

\begin{figure*}[tbp]
\centering
\includegraphics[width=0.95\linewidth]{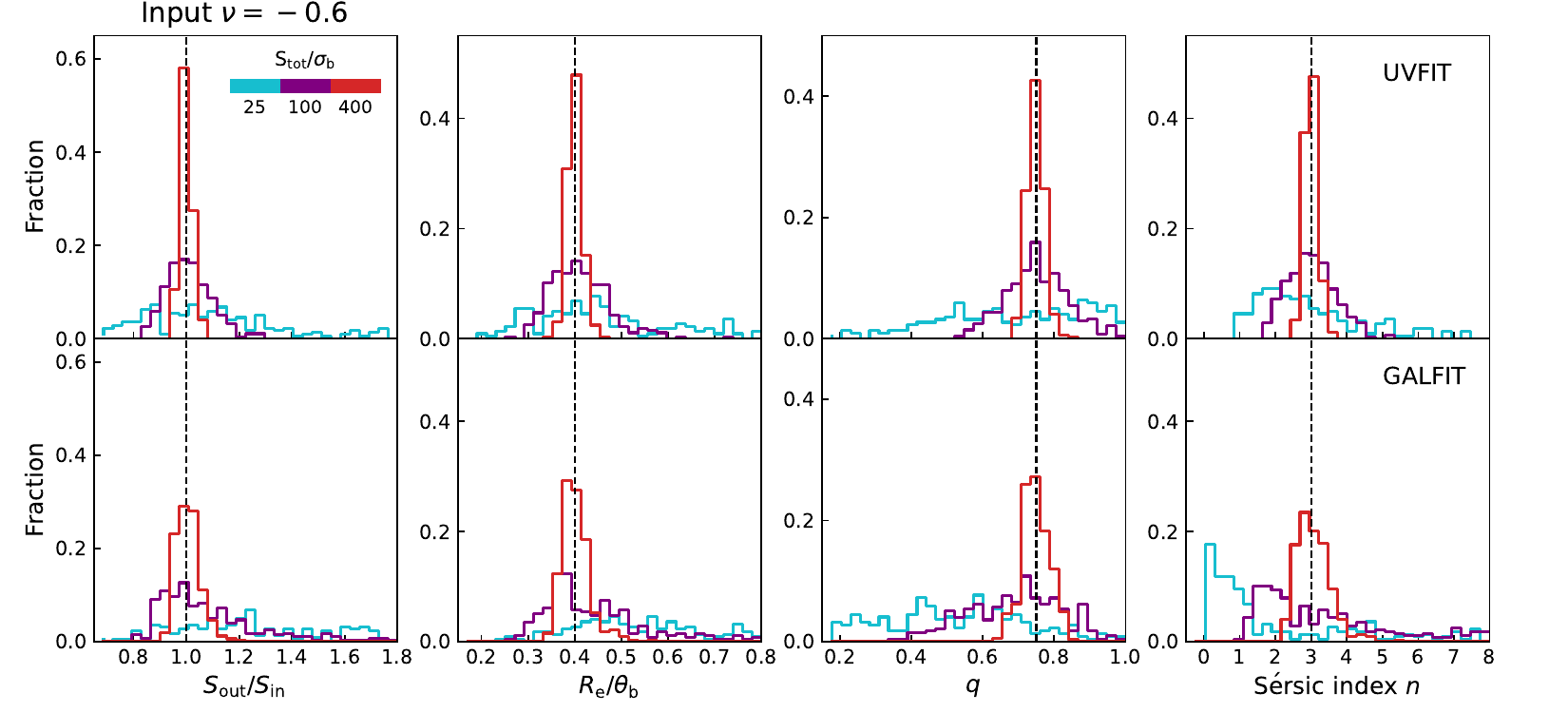}
\includegraphics[width=0.95\linewidth]{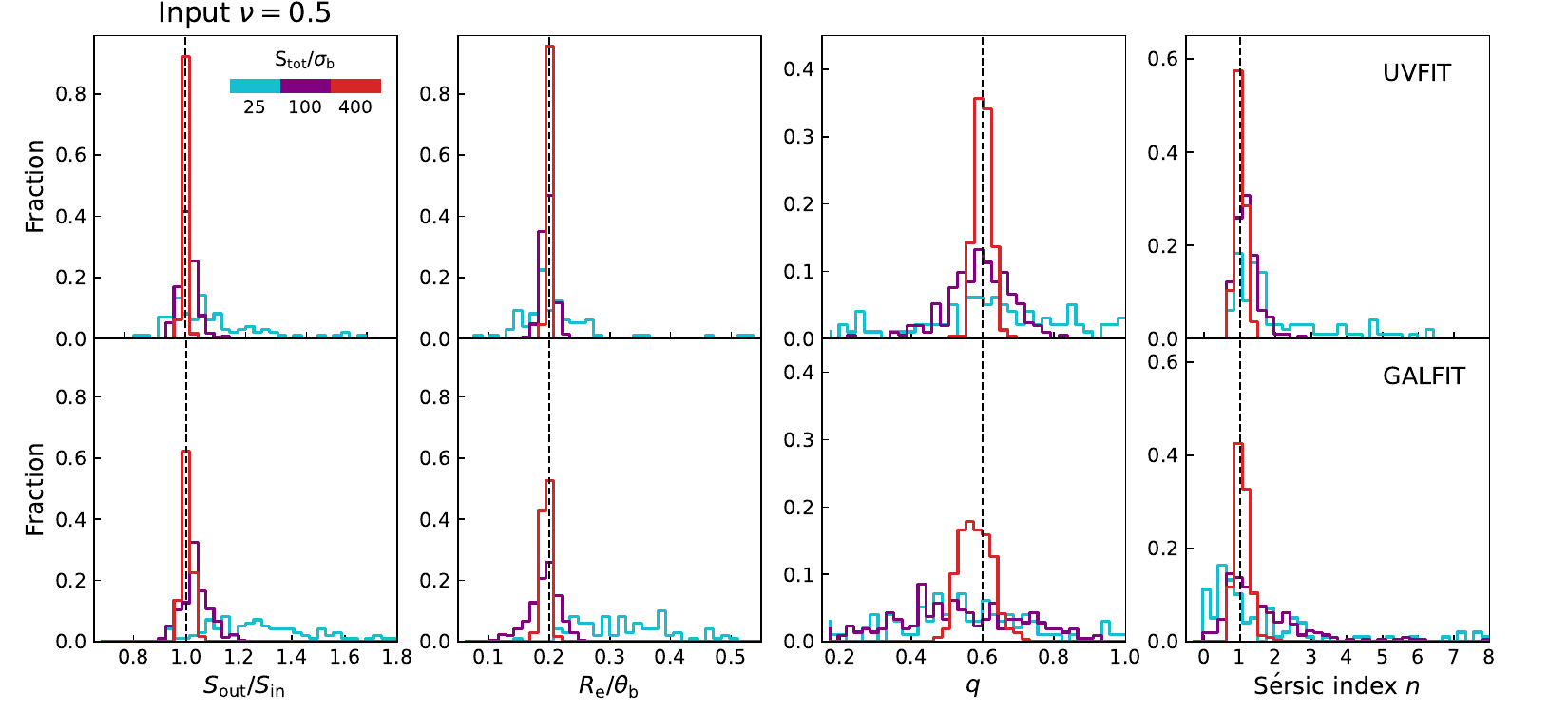}
\caption{Results from simulations of a single model source. The distribution of recovered parameters, from \texttt{uv$\_$fit} to the visibilities (panels labelled with \textsc{uvfit}) and {\it galfit} in the image plane (panels labelled with \textsc{galfit}), allows us to measure their respective accuracy in recovering the intrinsic known parameters (dashed vertical lines), including the flux density and structural parameters ($R_{\rm e}$, $q$, and $n$). We present two examples using a Spergel profile as input source: $R_{\rm e}/\theta_{\rm b}$ = 0.4, $q$ = 0.75, PA = 30$^\circ$, $\nu = -0.6$ (top-two panels), and $R_{\rm e}/\theta_{\rm b}$ = 0.2, $q$ = 0.6, PA = 30$^\circ$, $\nu = 0.5$ (bottom-two panels). The distribution of recovered parameter is color-coded by flux density ($S_{\rm tot}/\sigma_{\rm b}=$25, 100, and 400, indicated by different colors in the inset panel).  The \sersic \ indices shown in the \textsc{uvfit} panels are obtained by converting the best-fit Spergel indices, based on Eq.~(\ref{eq:conversion}). For all parameters, we have kept into account the conversion from Spergel-based fits to \sersic-based fits discussed in Section~\ref{sec:functions} (see Eqs.(\ref{eq:conversion}) and (\ref{eq:ratio-size-flux})), to remove any underlying systematics coming from the difference in the profiles.}\label{fig:hist}
\end{figure*}

\subsection{Details of the \texttt{UV\_FIT} implementation in \texttt{GILDAS}}\label{sec:uv-fit}

The \texttt{uv\_fit} command uses the \texttt{SLATEC/DNLS1E} implementation
of the Levenberg-Marquardt algorithm~\citep[see][for an intuitive
presentation]{press1992} to minimize the reduced $\chi^2$ of this non-linear
least-square problem. This algorithm only requires to deliver a routine
that computes the complex function and its partial derivatives with respect
to the different fitted parameters. Appendix~\ref{app:spergel:details} delivers the equations for the elliptical Spergel profile and its partial
derivatives. Once the minimum of the least-square problem is found, the
routine \texttt{SLATEC/DCOV} is called to compute the covariance matrix at
this minimum. The diagonal elements of this covariance matrix are the
$\pm1\sigma$ uncertainties on each fitted parameters.

In estimation theory, the Fisher matrix, \fisher, quantifies the amount of
information in the least-square problem. When the noise on the measurements
(the visibilities) is well modeled by an uncorrelated centered white
Gaussian random variable of standard deviation $\sigma$, the computation of
the Fisher matrix reduces to~\citep{stoica2005}
\begin{equation}
  \forall(i,j) \quad
  \bracket{\fisher}_{ij} = %
  \sum_{k=1}^{\nvisi{}} \frac{1}{\sigma_{k}^2} %
  \frac{\partial \visi_{k}}{\partial \varphi_i} %
  \frac{\partial \visi_{k}}{\partial \varphi_j},
  \label{eq:fisher:matrix}  
\end{equation}
where $[\fisher]_{ij}$ stands for the term $(i,j)$ of the Fisher matrix,
$\sigma_k$ and $V_k$ are the noise and the fitted visibility function for
visibility $k$, and $(\varphi_i)$ is the vector of fitted parameters. In our
case, $(\varphi_i) = (x_0,y_0,\Lo,\Rmaj,\Rmin,\PA,\nu)$, i.e., the central
position of the Spergel profile as an offset with respect to the phase
center, its luminosity, its major and minor half-light radius, its position
angle, and its index. The Cramer-Rao Bound (CRB) for each fitted parameter,
$\CRB(\varphi_i)$, is defined as the $i$th diagonal element of the inverse of
the Fisher matrix
\begin{equation}
  \CRB(\varphi_i)=\bracket{\fisher^{-1}}_{ii}.
\end{equation}
The CRB is the reference precision of the least-square problem. Indeed, the
variance of any unbiased estimator of the parameter $i$ will always be
larger than the associated CRB~\citep{garthwaite1995}, or
\begin{equation}
  \emr{var}(\varphi_i) \ge \CRB(\varphi_i).
\end{equation}
In other words, an efficient fitting algorithm will deliver variances for
the estimated parameters, which reach the associated CRB values. In
practice, sufficiently large signal-to-noise ratios are required to ensure
that the $\chi^2$ minimization converges towards the actual
solution. Additional explanations and an example of application to the fit
of CO(1-0) profiles in the local inter-stellar medium can be found
in~\citet{roueff2021}.

\section{Analysis: Comparison of $uv$-plane and image plane performances}\label{sec:comparison}

In this section, we compare the parameter estimates obtained by fitting Spergel profile in the $uv$-plane and \sersic \ profile in the image-plane to the same data, and thus comparatively assess their performances for the recovery of all the key parameters. 
Additionally, we investigate the reliability of the uncertainties returned by the fitting codes. For all parameters, the systematic terms that arise due to the intrinsic differences in the profiles (Section~\ref{sec:functions}, see Eqs.(\ref{eq:conversion}) and (\ref{eq:ratio-size-flux})) are always included in the comparison.

\subsection{Comparison of structural parameter measurements}\label{subsec:comp-structure}

In Fig.~\ref{fig:hist}, we present the results of our simulations, using only two models of galaxies as typical examples, for clarity. The input mock source was chosen with $R_{\rm e}/\theta_{\rm b}$ of 0.4 (0.2), axis ratio $q$ of 0.75 (0.6), and Spergel index $\nu$ of $-0.6$ (0.5). The distribution of recovered fitted parameter is shown in the top-two (bottom-two) panels. The $\nu=0.5$ case is particularly useful as it provides perfect coincidence with the Spergel and Sersic ($n=1$) models.

We extracted the distribution of the structural parameters obtained by fitting general Spergel profiles (panels labelled with \textsc{uvfit}) and \sersic \ profiles (panels labelled with \textsc{galfit}) with all parameters free, respectively (Fig.~\ref{fig:hist}).  The distribution of recovered best fitting values is shown for flux density, size, axis ratio, and \sersic \ index, where the true values are shown by vertical lines. The dispersion of recovered values can be used as a gauge of the measurement uncertainty. The average difference between recovered and true values probes any measurement biases and accuracy. 
In all cases, the scatter of the distributions decreases as the S/N of the simulated sources increases, as expected. Similarly, any systematic bias decreases with S/N, both in the image plane and in the $uv$-plane.

These two examples demonstrate that, when comparing fits in the $uv$-plane to measurements of each structural parameter obtained from profile fitting in the image-plane with {\it galfit}, it is clear that the latter exhibit larger scatter and have larger uncertainties.
Also, there is evidence for systematic biases that are more pronounced for image-plane fitting with {\it galfit}, especially at low S/N and clearly visible for both the low and high \sersic\ cases (albeit larger for the latter).

\begin{figure*}[tbp]
\centering
\includegraphics[width=1.0\linewidth]{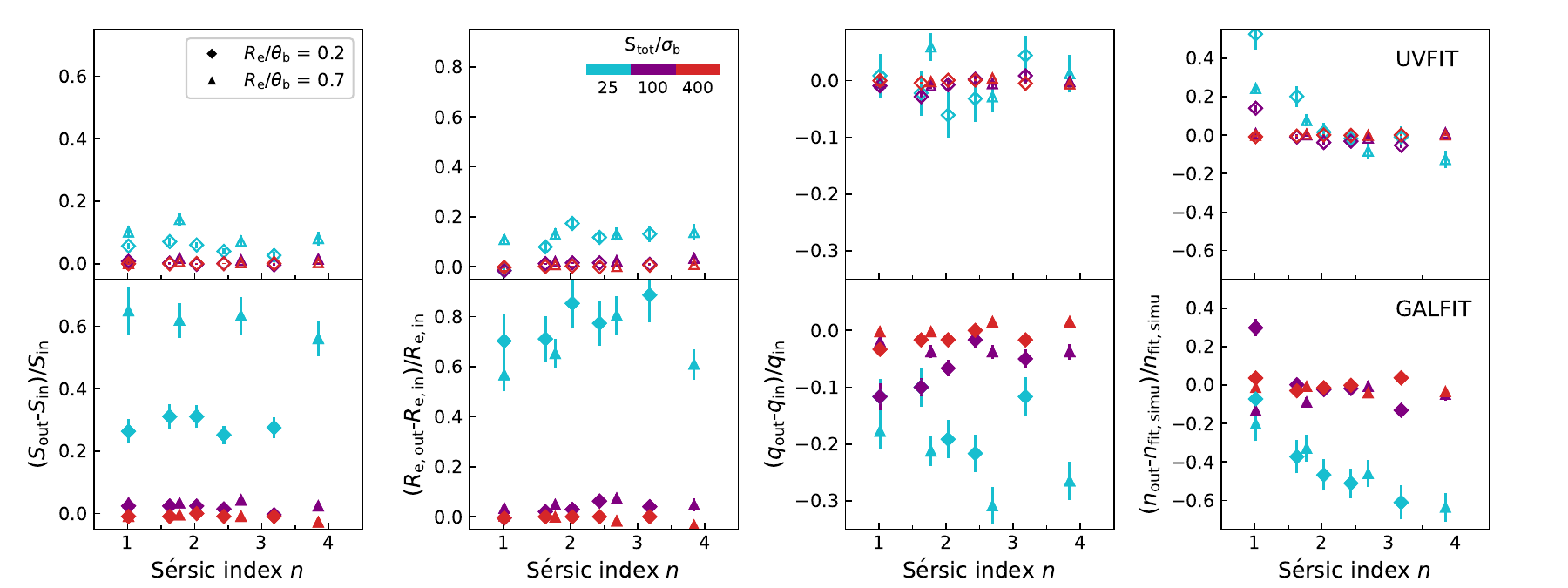}
\caption{Relative difference between the input and measured parameters, flux density, effective radius, axis ratio, and \sersic \ index (from left to right), obtained from Spergel fits to the $uv$-plane (top panels) and \sersic \ fits in the image-plane using {\it galfit} (bottom panels) in our simulations, plotted as a function of light concentration (i.e., S${\rm \acute{e}}$rsic index) for various $S_{\rm tot}/\sigma_{\rm b}$ and source sizes. Sources with a size of $R_{\rm e}/\theta_{\rm b}=$ 0.2 and 0.7 are marked by diamond and triangle respectively, while the inset panel of colorbar show the different input flux density. The error bars denote the interquartile range of the distribution divided by the square root of the number of simulations.
}
\label{fig:fit-summary}
\end{figure*}

The systematic relative deviations of all key parameters of the fit (biases) for a larger variety of input parameters are shown in Fig.~\ref{fig:fit-summary}, which again shows how the \sersic\ fits gets increasingly biased at low S/N much more rapidly than uv-plane fits. As the biases vanishes at the highest S/N even for the \sersic\ case, we conclude that these are not due to the previously discussed systematic differences in the profiles. 

Fig.~\ref{fig:error-ratio} compares the measurement uncertainty between $uv$-plane and image-plane. To enhance the readability of the figure, we plotted only the measurements obtained from fitting model sources with a size of $R_{\rm e}/\theta_{\rm b}$ = 0.2 and 0.7, and flux density of $S_{\rm tot}/\sigma_{\rm b}$ = 25, 100, and 400 in Figs.~\ref{fig:fit-summary} and~\ref{fig:error-ratio}. 
The following sections provide detailed results on the recovery bias of individual key parameters. Then, the focus shifts to the comparative uncertainty in parameters estimates. Again, we emphasize how the case $\nu=0.5$/$n=1$ is included, where the \sersic \ and Spergel models are identical, and it behaves fully similar to the other $\nu$/$n$ cases, demonstrating that the results are not driven by small differences between models at higher $n$.

\subsubsection{Flux measurements}

The left panels of Fig.~\ref{fig:fit-summary} show the median relative difference between the recovered and input flux densities, given by $(S_{\rm out}-S_{\rm in})/S_{\rm in}$, plotted against \sersic \ $n$ (converted from input $\nu$ using Eq.~(\ref{eq:conversion})). 
For both methods, there is a positive bias in recovering the flux density, leading to an overestimate of the flux. The magnitude of this bias depends on both the S/N  (S$_{\rm tot}$/$\sigma_{\rm b}$) and the source extension ($R_{\rm e}/\theta_{\rm b}$).

For image-plane fitting with {\it galfit} our simulations show that even relatively compact  sources with S$_{\rm tot}$/$\sigma_{\rm b}$ of 25 exhibit systematic errors of $>20$\% in the recovered flux densities. Large-sized sources with $R_{\rm e}/\theta_{\rm b}=0.7$ can be boosted by approximately $60\%$ of flux density at S/N$\sim25$. In contrast, the systematic biases of flux densities measured from $uv$-fitting are much smaller, with a mean value of $\sim$5\% at the faintest S/N$\sim25$.

\subsubsection{Size measurements}

 In terms of size estimates, the $uv$-plane method shows significantly smaller systematic offsets than image-plane measurements (see the second column panels of Fig.~\ref{fig:fit-summary}). Both methods have higher relative biases on $R_{\rm e}$ than flux density. In general, both methods tend to overestimate $R_{\rm e}$ for sources with low $S_{\rm tot}/\sigma_{\rm b}$. 

In the image-plane, 
for a compact, faint (S$_{\rm tot}$/$\sigma_{\rm b} = 25$) source with disk-like profile, the $R_{\rm e}$ can be overestimated by up to 70\%. On the other hand, the visibility-based $R_{\rm e}$ is on average overestimated by only about 15\% for the same model source.  

\subsubsection{Axis ratio measurements}

In Fig.~\ref{fig:fit-summary} (third column panels), the accuracy of recovering the axis ratio $q$ is compared by fitting in the image-plane and $uv$-plane, respectively. Both image-based and visibility-based measurements tend to systematically underestimate the recovered $q$ in most cases, but again the bias is much stronger in the image-plane. In addition, the accuracy of $q$ estimates is influenced by the size of galaxies, with the highest bias occurring for the most compact sources.

\subsubsection{S${\rm \acute{e}}$rsic index measurements}

For image-based measurements, the \sersic\ $n$ estimate is often biased towards a lower value in most cases (Fig.~\ref{fig:fit-summary}, fourth column). There is a significant increase in systematic offsets of measured \sersic \ $n$ as light concentration increases. In other words, the difference in estimating $n$ becomes large at low S/N and when the galaxy profile is steep. For the faintest source (i.e., S$_{\rm tot}$/$\sigma_{\rm b} = 25$) with a {\it de \ Vaucouleurs}-like profile in the simulation, the \sersic \ $n$ estimates can be underestimated by about 70\%. The underestimation goes down to 30\% when the source becomes less centrally concentrated with a disk-like profile. 

While the visibility-based concentration index shows systematic offsets at low S/N, it is still much less biased than the image-based one. The bulk of the error on $n$ estimates obtained from the $uv$-plane is limited to within about 20$\%$ or less. It should be noted that for very compact sources at the lowest S/N, the estimates of $n$ tend to be higher than the true value. Furthermore, the systematic error in the fit is greater for sources with an exponential profile than those with a steeper profile.

\begin{figure*}[tbp]
\centering
\includegraphics[width=\linewidth]{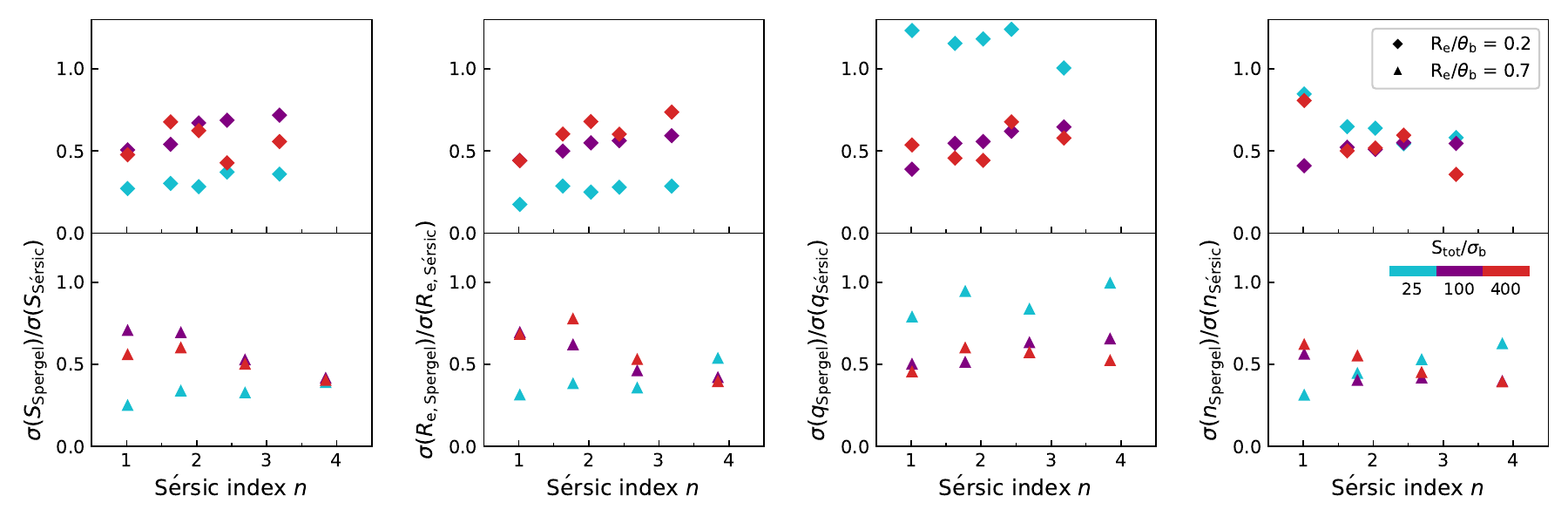}
\caption{Similar to Fig.~\ref{fig:fit-summary}, but we plot the ratio of the scatter of the measurements derived from Spergel fits in the $uv$-plane to that derived from \sersic \  fits in the image plane for fitted parameters. From left to right, we show flux density, effective radius, axis ratio, and \sersic \ index, as a function of light concentration (i.e., \sersic \ index). }\label{fig:error-ratio}
\end{figure*}

\subsection{Comparison of the relative measurement uncertainty}\label{subsubsect:scatter}

Apart from systematic biases, it is also important to evaluate the scatter of the measurements, both in the {\it uv}-plane and in the image plane, to verify if any of the two returns better measurements with lower scatter. 

In order to evaluate such relative measurement uncertainty, we use the scatter of the distribution of recovered values, evaluated as the median absolute deviation (MAD) of the data around the true value (i.e., input mock value) converted to  $\sigma$ using $\sigma=$1.48$\times$MAD, which is what expected for a Normal distribution. The use of MAD is preferred in order to to be less dependent on outliers while capturing the bulk of the spread in the sample. By defining the deviation with respect to the true value, the measurement bias is also taken into account when evaluating the overall uncertainty in the measurements. Fig.~\ref{fig:error-ratio} compares the scatter of measurements obtained from $uv$-fitting with those obtained in the image-plane for each key structural parameter.

Again, we find that the random uncertainties in all structure parameter recoveries are systematically larger for image-based estimates than those obtained by visibility analysis. The median scatter ratio of measurements obtained from $uv$-fitting to that obtained from image-plane for the recovered flux densities is 0.5, while for the recovered effective radius, axis ratio, and \sersic \ $n$, the median scatter ratios are 0.5, 0.6, and 0.5, respectively. We did not find a significant correlation between the scatter ratio and the S/N of data. These findings are in line with a study that compared the performance of stacking data in the $uv$-plane and image-plane, which found that $uv$-stacking resulted in significantly improved accuracy of size estimates, with typical errors less than half compared to image-stacking \citep{lindroos15}. We emphasize that the small residual systematics shown in Figs.~\ref{fig:index-matching} and \ref{fig:infinite-simu}, inherent in the conversion of Spergel-based parameters of our simulations to \sersic \ parameters, only account for a negligible amount of the excess noise coming from {\it galfit} (and have no effect at all for the $n=1$ case). 

In conclusion, we find that measurements in the image-plane are not only more subject to bias, but also returning parameters that are more substantially affected by noise, compared to those in the $uv$-plane.

\section{Analysis: absolute accuracy of $uv$-plane modeling and reliability of uncertainties}\label{sec:spergel_abs}

In this section, we focus on the performances of the Spergel model fitting in GILDAS \texttt{uv$\_$fit}. We analyze its accuracy in retrieving intrinsic galaxy morphological parameters and verify the reliability of parameter errors returned by the code. In addition, we evaluate the presence of correlations between fitted parameters in the presence of noise.

\begin{figure*}[tbp]
\centering
\includegraphics[width=0.95\linewidth]{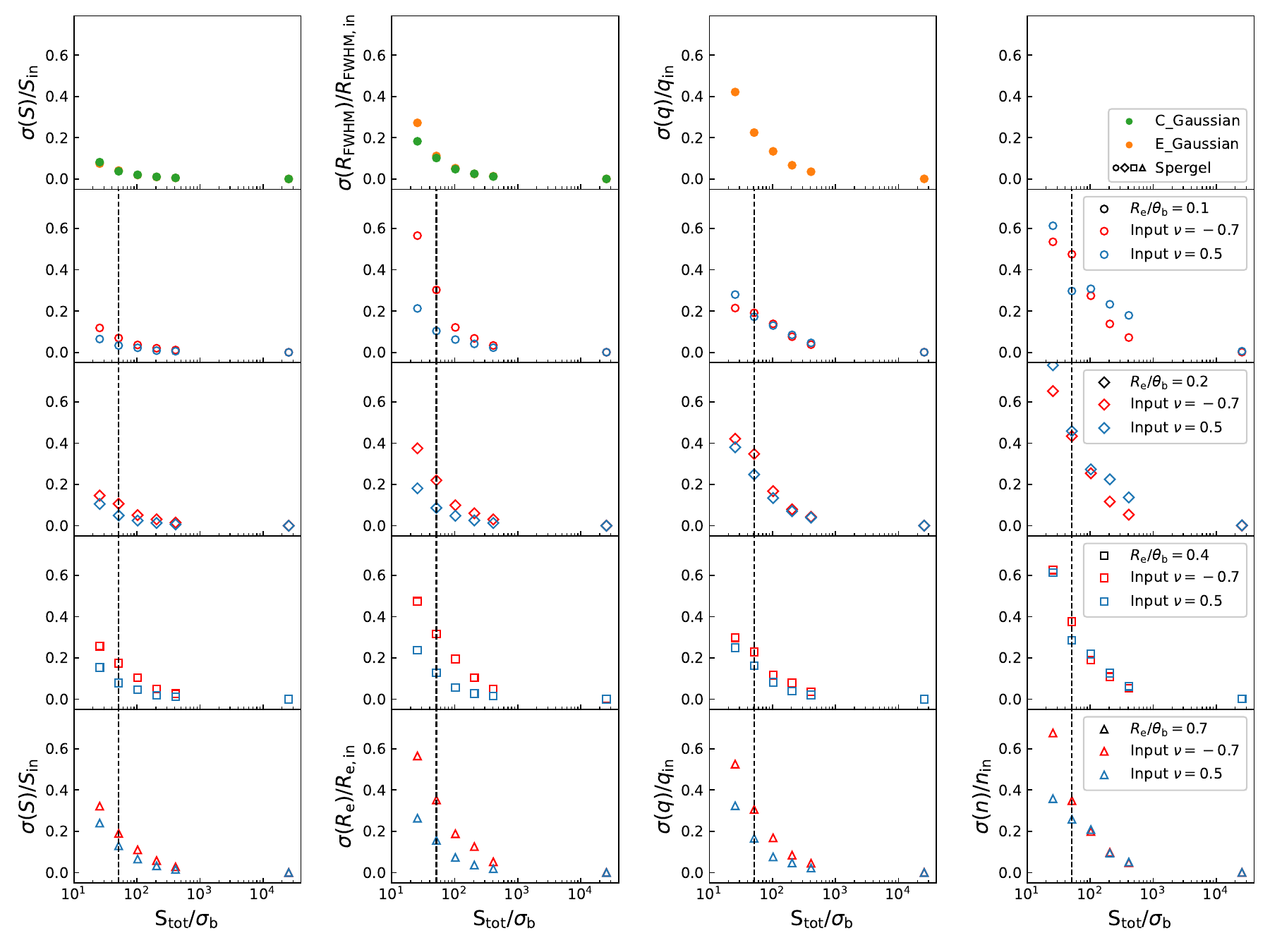}
\caption{The relative accuracy of the $uv$-based parameter estimates, flux density, effective radius, axis ratio, and \sersic \ index (from left to right), given by $\sigma$(para)/para, where $\sigma$(para) is the uncertainties of the parameter estimates distribution and calculated as 1.48$\times$MAD, as a function of S/N of flux density. The top panels show the results derived from fits with an elliptical Gaussian (orange) and circular Gaussian (green) model, while the results from a Spergel model $uv$-fit with input $R_{\rm e}/\theta_{\rm b}$ of 0.1, 0.2, 0.4, and 0.7 are shown in panels from second to fifth rows, respectively. Here we only display results for the cases with a Spergel index of 0.5 and $-0.7$ in our simulation. For other cases with a Spergel index between $\nu=0.5$ and $-0.7$, the results are found to be within or close to those shown in panels between the second and fifth rows. The dashed vertical lines represent the threshold of $S_{\rm tot}/\sigma_{\rm b}$ of 50.
}\label{fig:fit-accuracy}
\end{figure*}

\subsection{S/N recoverable for fitted parameters in the $uv$-plane}
 
We check the achievable  accuracy for the  parameter estimates from Spergel fits in the uv-plane by computing the ratio of the uncertainties of the parameter estimates to the input mock value, given by $\sigma$(para)/para. The uncertainties of the parameter estimates, $\sigma$(para), can be evaluated as 1.48$\times$MAD, as discussed in Section~\ref{subsubsect:scatter}.

In Fig.~\ref{fig:fit-accuracy}, the accuracy of parameter estimates is shown to vary with the S/N of the data. The parameters being estimated are flux density, size, axis ratio, and \sersic \ index. To enhance readability of the figure, only Spergel models with an exponential ($\nu = 0.5$) profile and steep profile with $\nu=-0.7$ (close to a {\it de Vaucouleurs} profile) were considered. These models are shown in the panels between the second and fifth rows. It is worth noting that for other cases where the Spergel index is assumed to be between $\nu=0.5$ and $-0.7$, the measurements are found to be either within or close to the values obtained from the above two cases. Therefore, the results presented in Fig.~\ref{fig:fit-accuracy} can be considered representative of the Spergel model.

Fig.~\ref{fig:fit-accuracy} shows that the uncertainties in measuring galaxy shape parameters, such as size and Spergel index, are significantly larger than those for  flux density. The most difficult parameter to estimate is the Spergel index. The uncertainty in estimating the Spergel index can be as high as $70\%$ for model sources with a S$_{\rm tot}$/$\sigma_{\rm b}$ of 25 or lower, suggesting that the measurement of Spergel index can be highly uncertain. As the S$_{\rm tot}$/$\sigma_{\rm b}$ increases to 50, the accuracy of Spergel index estimates significantly improves, with a median value of $\sigma$($n$)/$n$ of 0.36, which we consider as the bare  minimum to define a meaningful estimate. At S$_{\rm tot}$/$\sigma_{\rm b}$ of 50, the uncertainties of the flux density, size, and axis ratio are significantly smaller, with median values of 9\%, 18\%, and 22\% of estimates, respectively. To obtain a meaningful and reliable profile fitting result with the Spergel model in $uv$-plane, our simulations strongly suggest that a S$_{\rm tot}$/$\sigma_{\rm b}$ of at least 50 should be required.

In addition, we find that the accuracy of both the flux density and size is lower for simulated data with a steep profile compared to the model data with a flat profile. On average, the accuracy of flux densities and sizes for the galaxy with an exponential profile ($\nu=0.5$) is 50$\%$ higher than those of the galaxy with a {\it de Vaucouleurs}-like profile ($\nu=-0.7$). However, it is important to note that the measurements of Spergel $\nu$ tend to be less accurate for smaller sources ($R_{\rm e}/\theta_{\rm b}\leqslant$0.2) compared with larger sources in the data. We have also found that both the flux densities and sizes of the small-sized sources are more accurately measured, with a typical factor of about 40\%. The differences in the accuracy of parameter estimates imply that all the fitted parameters are interrelated. 

\begin{figure*}[tbp]
\centering
\includegraphics[width=0.25\linewidth]{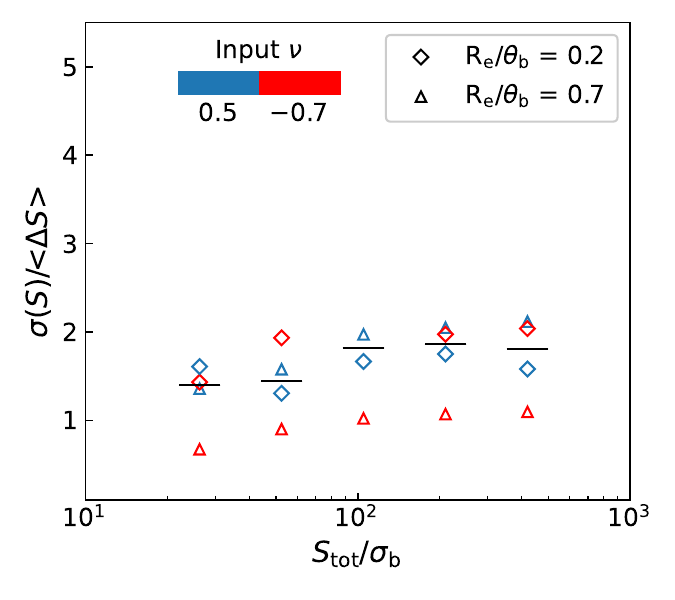}
\hspace{-10pt}
\includegraphics[width=0.25\linewidth]{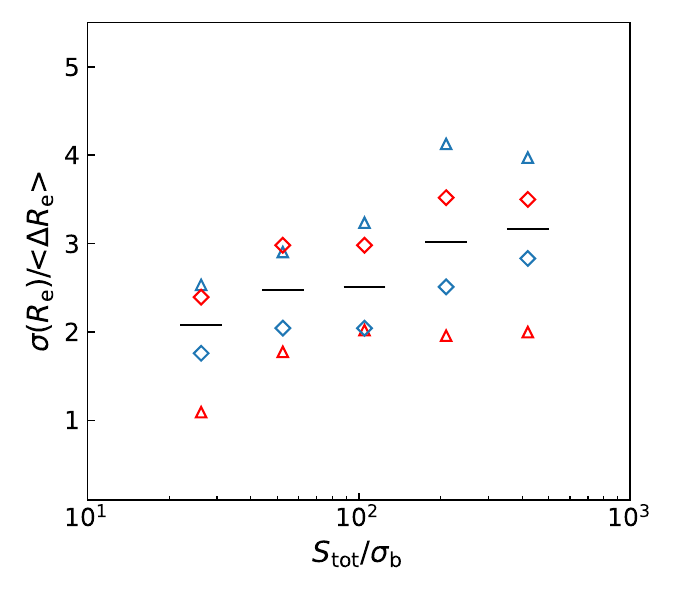}
\hspace{-10pt}
\includegraphics[width=0.25\linewidth]{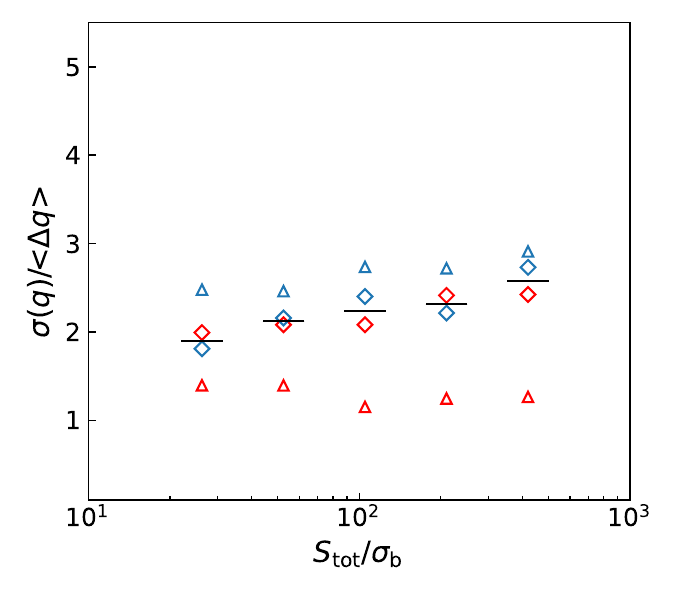}
\hspace{-10pt}
\includegraphics[width=0.25\linewidth]{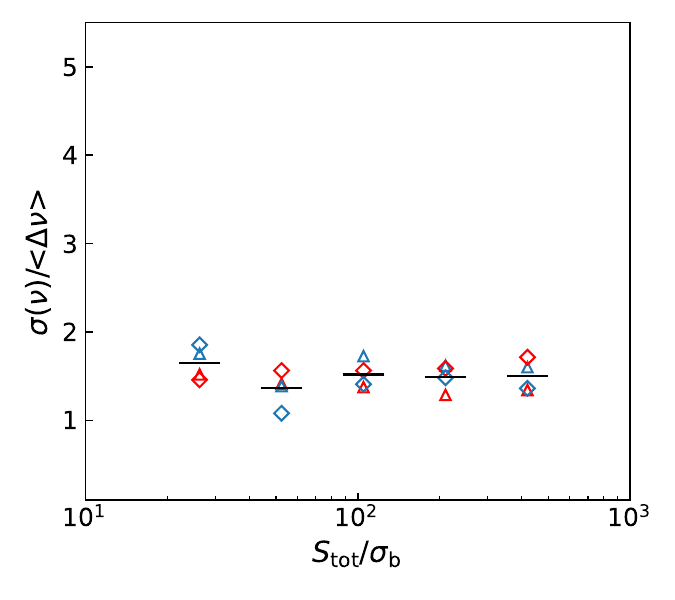}
\caption{Comparison between the errors obtained from the scatter of the distribution of parameter estimates and the errors estimated by \texttt{uv$\_$fit} at each fit in our simulation, as a function of the S/N of flux density. The fitted parameters, from left to right, are flux density, effective radius, axis ratio, and Spergel index. The scatter of the parameter estimates distribution is calculated as $\sigma$ = 1.48$\times$MAD. Simulated sources with a size of $R_{\rm e}/\theta_{\rm b}=$ 0.2 (diamond) and 0.7 (triangle) are presented and color-coded by the Spergel $\nu$. The black bars represent the median value of the ratio between the error in each fitted parameter obtained from simulations and the median of the error measured by \texttt{uv$\_$fit} at each flux S/N bin for the simulated sources in this work.}\label{fig:errors}
\end{figure*}


\subsection{Reliability of parameter uncertainties from UVFIT}\label{subsec:uncertainty_uvfit}

In Fig.~\ref{fig:errors}, we compare the average value of parameter errors in simulated galaxies to the posterior scatter of the recovered distributions to explore their reliability. We find that in most cases, the uncertainties of parameters measured from Spergel profile fitting are underestimated, although typically within a factor of 2. The underestimation is generally less important for steep profiles ($\nu= -0.7$) than for disks ($\nu=0.5$) and for extended versus compact galaxies. At each flux S/N bin, the ratio between the error obtained from simulations and the median of the error measured by \texttt{uv$\_$fit} for the whole set of simulated sources is in the range of 1.4--1.9, 2.0--3.2, 1.9--2.5, and 1.4--1.6 (see the black bars in Fig.~\ref{fig:errors}), with a mean value of 1.6, 2.5, 2.1, and 1.5 for the parameter estimates of flux density, effective radius, axis ratio, and Spergel index, respectively.

\begin{figure*}[tbp]
\centering
\includegraphics[width=0.9\linewidth]{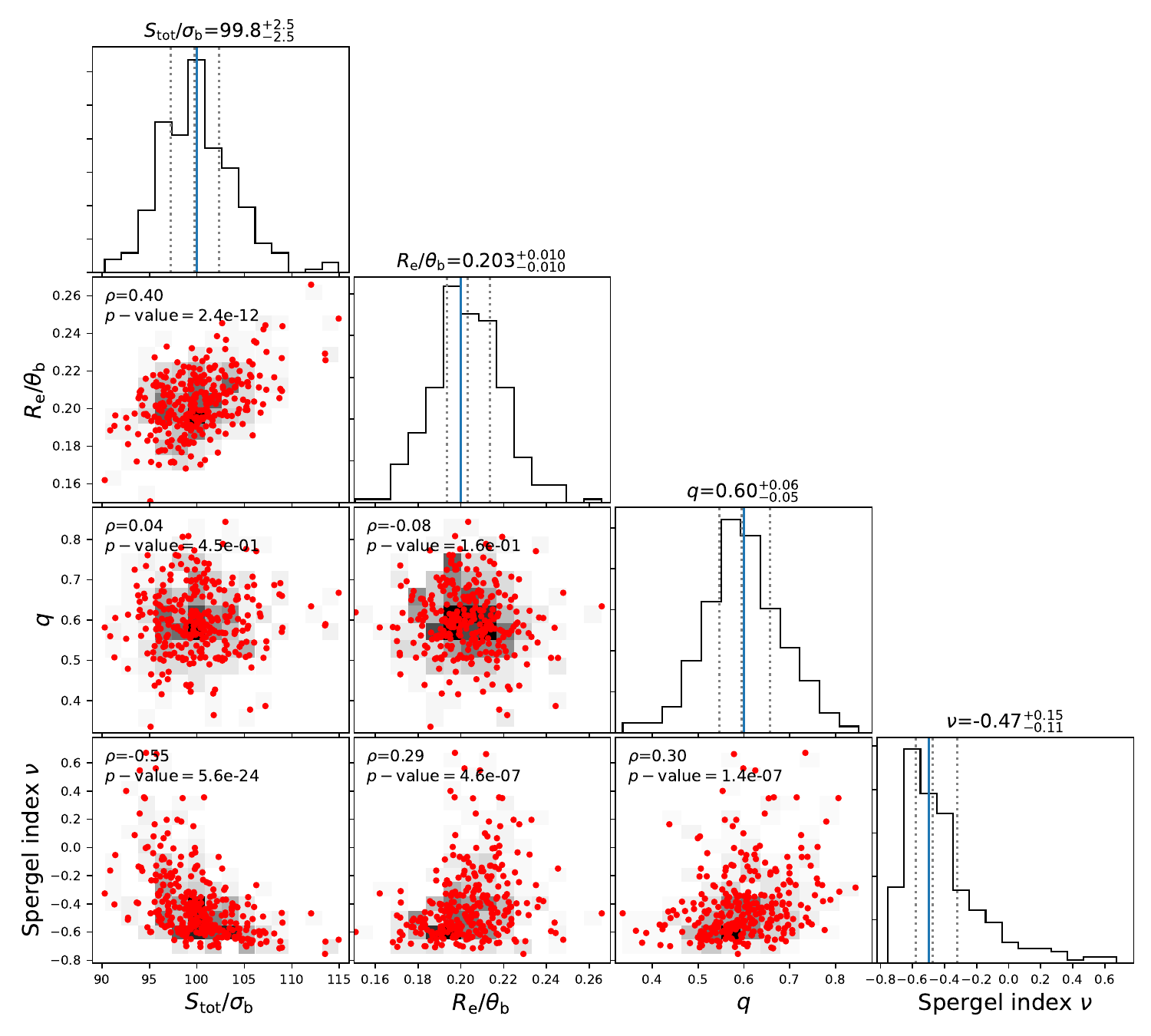}\\
\includegraphics[width=0.95\linewidth]{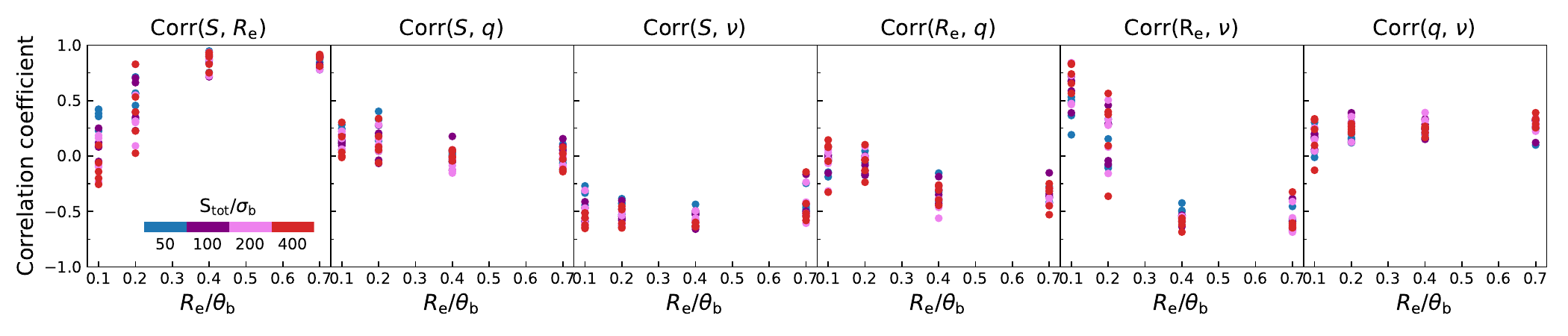}\\
\caption{{\bf Top}: corner plots showing the covariances between the free parameters modeled in Spergel profile fitting. The example model source has a flux density of $S_{\rm tot}/\sigma_{\rm b}=100$, a size of $R_{\rm e}/\theta_{\rm b}=0.2$, an axis ratio of $q$=0.6, and a Spergel index of $\nu=-0.5$. The shaded density histograms show the two-parameter distributions with Spearman rank correlation coefficient and p-value marked in each panel. The one-dimensional histograms at the top of each column represent the individual parameter distributions, annotated with the median values. The boundaries of the 25th and 75th percentiles of the distribution are plotted as dashed lines, while blue vertical lines show the true values. {\bf Bottom}: Spearman's rank correlation coefficients of the two-parameter pairs as a function of source size. The points are colour-coded by the input flux density.}\label{fig:covar}
\end{figure*}

\subsection{Covariance of Spergel model fitted parameters}

In this section, we examine the covariance between fitted parameters estimated from \texttt{uv$\_$fit} using a Spergel model. The top panels in Fig.~\ref{fig:covar} show the correlations between the fitted parameters, which are flux density $S_{\rm tot}/\sigma_{\rm b}$, effective radius $R_{\rm e}/\theta_{\rm b}$, axis ratio $q$, and Spergel index $\nu$, for a simulated dataset with an input $S_{\rm tot}/\sigma_{\rm b}$ of 100, $R_{\rm e}/\theta_{\rm b}$ of 0.2, $q$ of 0.6, and $\nu$ of $-0.5$, respectively. In this case, the Spearman correlation coefficient, which is calculated by dividing the covariance by the intrinsic scatter of each parameter, show weak correlations ($\vert \rho \vert \lesssim 0.3$) between fitted parameters of size, axis ratio, and $\nu$, while moderate correlations ($\vert \rho \vert \sim 0.5$) are found between the flux density and both $\nu$ and size. 

The bottom part of Fig.~\ref{fig:covar} summarizes the pairwise correlation coefficients calculated for all the data sets in our simulation. Generally, we find that the correlation between variables becomes more prominent as the S/N increases, except for the correlation between flux density and size. For sources that are significantly more compact than the beam ($R_{\rm e}/\theta_{\rm b}<0.2$), the correlation between the flux density and measured source size is weaker when the source is detected with a high S/N. We did not find any significant correlation between {\it q} and other parameters. This indicates that the measured axis ratios are almost independent of these parameters.

There is a positive correlation between flux density and \sersic \ $n$ (anti-correlated with Spergel $\nu$), indicating that sources with higher measured \sersic \ $n$ tend to have a larger flux density. In addition, a strong positive correlation is seen between flux density and size, except for very compact sources with $R_{\rm e}/\theta_{\rm b}$ of 0.1, where the correlation is relatively weak. This means that sources for which sizes are overestmated  have a tendency to be also boosted in flux density. 

\section{Discussion}
\label{sec:discussion}

The results presented in the previous sections demonstrate that studying galaxies morphologies in the $uv$-plane leads to better performance than imaging the data and then using {\it galfit} to study morphologies in the image-plane. This approach offers exciting possibilities for studying morphologies of galaxies in SFR tracers. However,
there are several issues that merit discussion.

\subsection{On the differences between Spergel and \sersic\ profiles}\label{subsec:discuss-profiles}

A comparison between the Spergel and \sersic \ profiles shows that the Spergel model has a steeper core and faster declining wings compared to the \sersic \ model (see Fig.~\ref{fig:profiles} and Fig.~\ref{fig:app-profile-mcmc}), except for the case of Spergel $\nu=$0.5 which is equivalent to an exponential profile (\sersic \ $n$=1). The question of whether Spergel or \sersic \ models provide better fits to actual galaxies remains open and requires future investigation. The current available data quality may not be sufficient to determine a clear preference between the two functional forms, in absolute terms. For the time being and with the typical data available, we deem the two functional forms as equivalent. 

Converting a Spergel index to a \sersic\ one  (as well as $R_{\rm e}$ and total flux) requires knowledge of the intrinsic size of the galaxy and the angular resolution of the observations. In the case of this study, the angular resolution is determined by the synthesized beam of the interferometric data. We believe this requirement also applies implicitly to optical observations. When a \sersic \ index is derived from optical data, it applies to the range of scales actually observed in the data. It may not apply beyond the observed scales by definition. It is possible that a somewhat different Spergel/\sersic\ index might be recovered when re-observing the same galaxy with a much different surface brightness sensitivity. 

It is also relevant to question whether the differences in the profiles at the outer and inner ranges  could affect the systematic biases of the parameter estimates, particular for a steep profile. 
Along these lines, considering that we have simulated Spergel models and then fitted them with {\it galfit} in the image-plane, one might wonder if the under-performance of {\it galfit} in the image-plane could be simply related to the discussed differences between Spergel and \sersic\ models.

As already mentioned through-out the description of results in the previous section, 
we believe that the argument can be dismissed, but it's worth recalling the key evidences here. Based on the bottom panels of Fig.\ref{fig:fit-summary}, 
the simulations with the highest S/N (400 in this case) show that any average bias in \sersic\ modeling is vanishing or strongly reduced, as expected by construction (Eq.(\ref{eq:conversion})), showing that any residual systematics beyond our conversions do not have a measurable impact. The systematic differences in the recovered profile parameters ($R_{\rm e}$ and \sersic\ $n$, crucially) are in fact strongly S/N-dependent, and therefore mostly an effect of noise, rather than arising from structural differences. Additionally, Fig.\ref{fig:error-ratio} shows that the excess noise in parameter recovery from image-plane fitting is not a function of S/N, while the systematic deviations between the profiles are expected to become more evident at the highest S/N. 
Also importantly, Fig.\ref{fig:error-ratio} shows that an entirely consistent situation is seen for the $n=1$ case, which is fully identical in shape to a Spergel $\nu$ of 0.5.

\begin{figure*}[tbp]
\centering
\includegraphics[width=\linewidth]{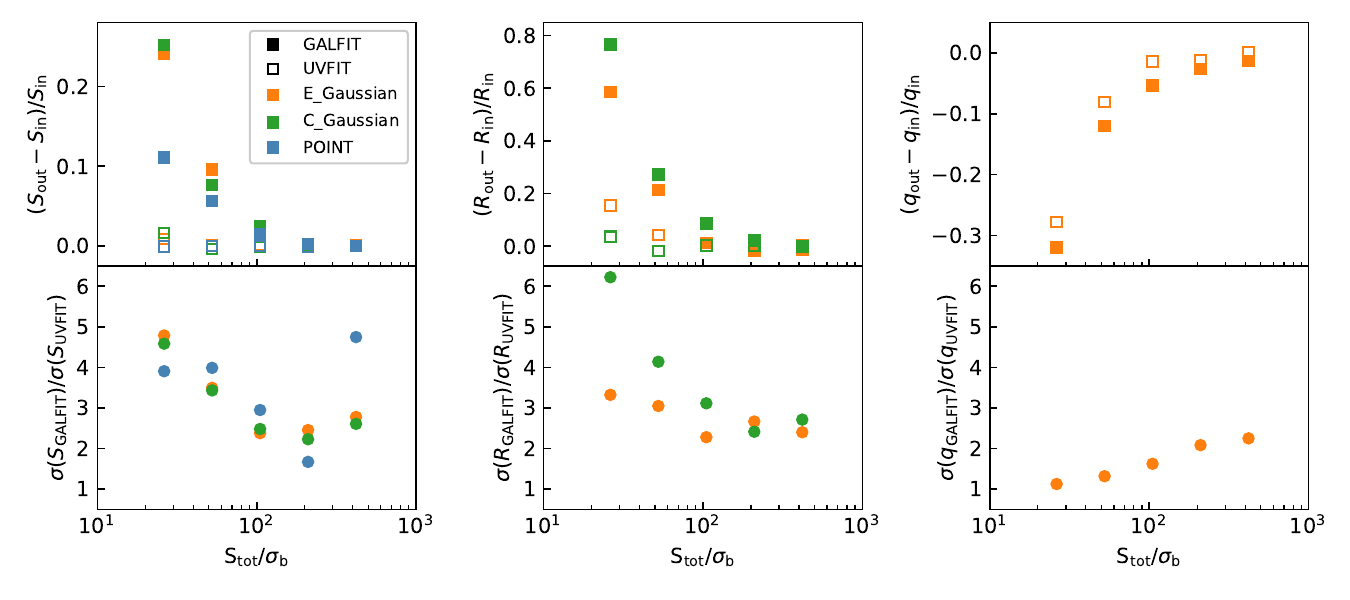}
\caption{{\bf Top}: relative difference between recovered and input parameters of flux density (left), size (middle), and axis ratio (right), which were obtained from fitting with an elliptical Gaussian (orange), a circular Gaussian (green), and a point (blue) source model using \texttt{uv$\_$fit} (open symbols) and {\it galfit} (solid symbols). {\bf Bottom}: similar to the top panels, but we plot the  ratio of measurement uncertainty derived from {\it galfit} in the image-plane to that obtained from \texttt{uv$\_$fit} in the $uv$-plane. The measurement uncertainties are estimated as $\sigma = $ 1.48$\times$MAD.}\label{fig:fit-gauss}
\end{figure*}

Fig.\ref{fig:fit-gauss} extends further this point by showing the performance comparison of  \texttt{uv$\_$fit} versus {\it galfit} in the case of Gaussian sources and PSF models. In this case, we directly simulated these shapes in the $uv$-plane, which are not Spergel models, and there is no shape difference with respect to the model fitted by {\it galfit}.
However, the overall behaviour is qualitatively the same as in Figs.\ref{fig:fit-summary} and~\ref{fig:error-ratio}. There are stronger biases in the image plane (top panels), and the effective noise is also much higher there (lower panels).

\subsection{On the origin of the poor performances in the image plane with GALFIT on interferometric data}
 
It has been already shown that ignoring noise correlation in interferometric images can lead to a significant underestimation of the statistical uncertainty of the results \citep[e.g.,][]{tsukui23}. To test this idea further, we carried out aperture photometry as an alternative to {\it galfit} for point-source simulations. Aperture photometry is one of techniques used to measure flux density in astronomical observations, which is also used in interferometric images  \citep[e.g.,][]{lang19,gomez-guijarro22a}. 
It should be noted that, for extended sources, there is an added complication of correcting for flux losses, and this technique has the limitation of not allowing to estimate morphological parameters. However, it is well-defined   for flux density measurements of point sources. 

\begin{figure*}[tbp]
\centering
\includegraphics[width=0.8\linewidth]{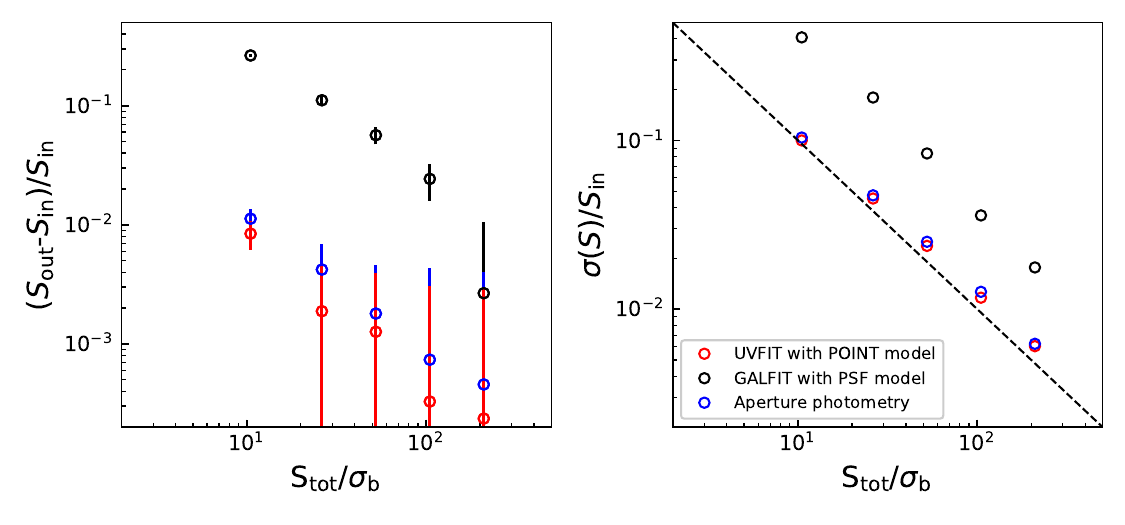}
\caption{Simulation results of the flux density bias (left) and the relative uncertainty of measurement on recovery of flux density (right) for point sources using \texttt{uv$\_$fit} (red), {\it galfit} (black), and aperture photometry (blue) in the image-plane fitting, as a function of the flux S/N. The error bars in the left panel represent the standard error on the mean. The dashed line in the right panel is a 1:-1 line.}\label{fig:aper}
\end{figure*}

Figure~\ref{fig:aper} shows the systematic bias and the measurement uncertainty on the recovery of the flux density for point sources using different fitting methodologies, i.e., \texttt{uv$\_$fit} with a Point source model, {\it galfit} with a PSF profile that is identical to the dirty beam of the simulated data, and aperture photometry. The latter is performed with an aperture radius equal to the circularized radius of beam size and at the position returned by {\it galfit} PSF fitting. It is clear that the flux density estimates obtained through aperture photometry are fully consistent  to those measured in the $uv$-plane, with a similar level of accuracy in the estimates. Both of these methods exhibit smaller systematic errors and scatters when compared to fitting with a PSF model using {\it galfit}. In a way, aperture photometry measurements being fairly basic and raw just are not fooled by false coherence in the signal induced by correlation, and return the full S/N performances in the limiting case of point-like emission (Fig.\ref{fig:aper}-right). 

\subsection{Warnings against {\it  cleaning} dataset in view of morphological measurements}

It is common practice to perform deconvolution of low-frequency radio interferometric data to remove strong sidelobes from bright sources in  very large fields, a process known as {\it cleaning}. We emphasize that in  ALMA and NOEMA observations the field of view is tiny compared to the radio and the source density is basically always manageable, so that {\it cleaning} can generally be safely avoided. Cleaning, as any deconvolution approach, is a model-dependent process which in most implementations arbitrarily interpolates the observed source with a superposition of point sources, whose side-lobes are then subtracted from the data using the {\it dirty beam}. Due to its arbitrariness and its approximation of real sources as multiple point-sources, it is clearly not ideal for analysis of galaxies morphological profiles. We finally emphasize that, obviously, the cleaning process would not remove correlations in the signal in the image plane, which are due to Fourier transforming the visibilities.

\begin{figure}[tbp]
\centering
\includegraphics[width=\linewidth]{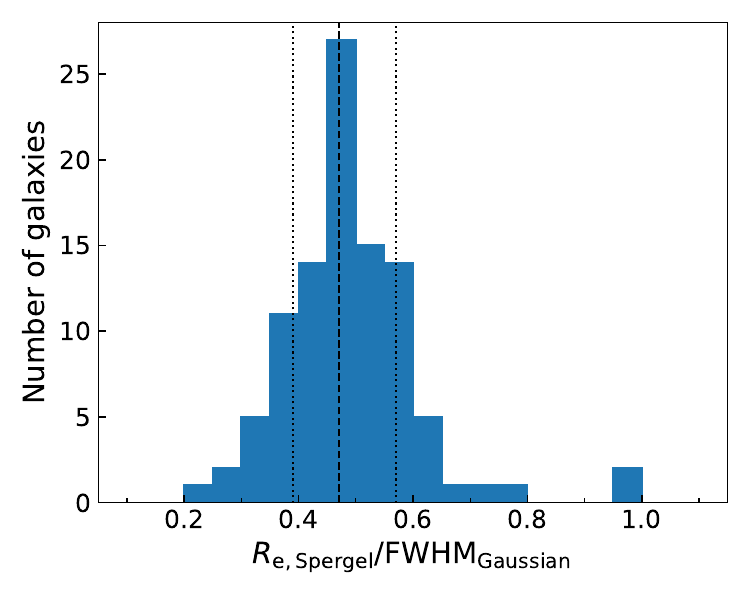}
\caption{Distribution of the ratio of the size obtained from a Spergel fit to that from a Gaussian fit. The dashed line shows the median of the distribution, while the dotted lines represent the 16$^{\rm th}$ and 84$^{\rm th}$ percentiles of the distribution, respectively. }\label{fig:size-gaussian-spergel}
\end{figure}

\begin{figure*}[tbp]
\centering
\includegraphics[width=0.8\linewidth]{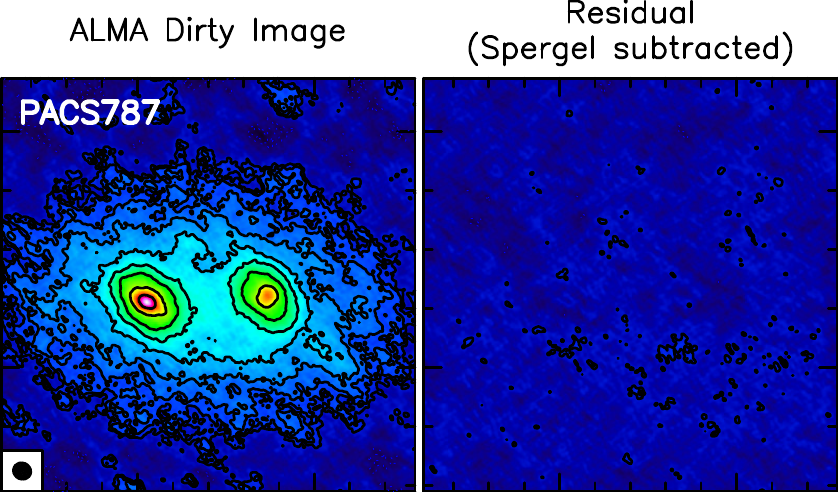}
\includegraphics[width=0.8\linewidth]{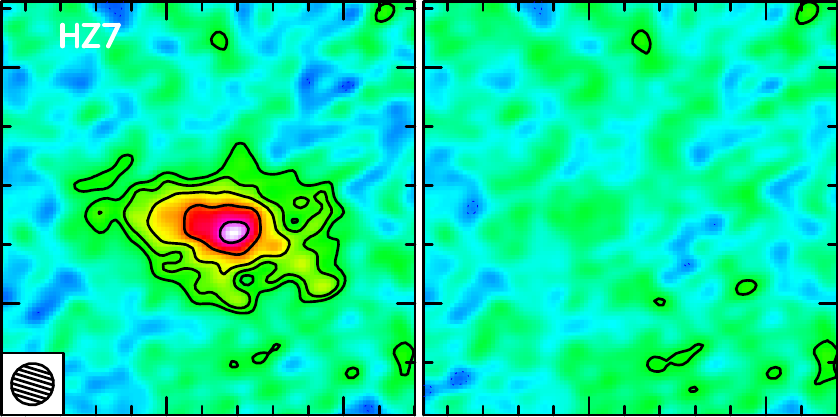}
\caption{Examples of ALMA images of CO(5-4) and [CII] emission from high-z star-forming galaxies PACS-787 (top) and HZ7 (bottom). Left: the ALMA dirty image. Right: the residual map with the primary disk component modeled by a single Spergel profile subtracted. The image size is $3.5\arcsec \times 3.5\arcsec$. The synthesized beam ($0.15\arcsec \times 0.14\arcsec$ and $0.36\arcsec \times 0.35\arcsec$ for PACS-787 and HZ7, respectively) is presented at the bottom left in the left panels. Contours start from $\pm 2\sigma$ and increase by a factor of 1.5.}\label{fig:real-data}
\end{figure*}

\subsection{The cost of using Spergel: recommendations}

To investigate the morphologies of galaxies observed in interferometric images, Spergel modeling directly in the $uv$-plane should be the preferred approach. However, a high S/N ratio of at least 50, and ideally much more, must be required to perform a Spergel fit that meaningfully constrain the Spergel index, as we have previously commented based on Fig.\ref{fig:fit-accuracy}. This requirement is
similar to optical observations of galaxies and their {\it galfit} modeling, where a high S/N of order 100 is needed to attempt derivation of a \sersic\ index \citep{vanderwel14,magnelli2023}. 

These considerations should be extended further to include other parameters of interest, such as the size or even the flux density. These parameters are easier to derive and require less S/N compared to a Spergel/\sersic\ index. However, there is a tension between extracting a meaningful measure from the data and ensuring an unbiased measure. It is important to determine the best approach in balancing these factors.

The top-panel of Fig.\ref{fig:fit-accuracy} provides information on the uncertainties in flux density, size, and axis ratio for Gaussian models, as a function of their S/N ratios. By comparing this information to the bottom panels, one can make a decision between a less complex (Gaussian) fit and a more complex (Spergel) fit to a given dataset. For example, Gaussian fits can provide size uncertainties of about 20\%  down to an S/N ratio ($S_{\rm tot}/\sigma_{\rm b}$) of 20, whereas Spergel fits do not offer the same level of accuracy at lower S/N ratios. Similar considerations apply to flux density and axis ratio. In addition, there are no significant differences in the accuracy and uncertainty estimates between circular and elliptical Gaussian models (see Fig.~\ref{fig:fit-accuracy} and Fig.~\ref{fig:fit-gauss}). This demonstrates that the increase in degrees of freedom in Spergel models leads to greater uncertainty in the measured parameters, including flux density and size. 

Figure~\ref{fig:size-gaussian-spergel} shows the distribution of the ratio of the size obtained from a Spergel fit to that from a Gaussian fit, measured for a sample of about 100 high-$z$ star-forming galaxies observed with the ALMA (Q. Tan et al., in preparation). The median ratio between the $R_{\rm e}$ size obtained from a Spergel fit and the FWHM size obtained from a Gaussian fit is $0.47_{-0.08}^{+0.10}$, where uncertainties correspond to the 16th and 84th percentiles of the distribution. This implies that in most cases, the FWHM size measured from a Gaussian model can be used as a good approximate of the effective radius using the relation $R_{\rm e}$=FWHM/2. However, it is worth noting that almost all of the outliers with $R_{\rm e}$/FWHM ratio far from the median value in Fig.~\ref{fig:size-gaussian-spergel} are sources measured with large \sersic \ index $n$ ($n>2$; Q. Tan et al., in preparation). This suggests that the difference in size measured from Gaussian fit and the Spergel fit could be significant when the source has a large \sersic\ index.

Finally, to achieve a global optimal solution for the Spergel fit, we recommend utilizing the fitting results obtained from Gaussian or exponential fits as prior knowledge for the initial guesses of each structure parameter.  This provides good starting guesses, which helps the solver in achieving convergence and in providing a reasonable level of uncertainty. Such preliminary information will also inform the user about the actual merit of proceeding to a more demanding full-Spergel fit. 

\subsection{The power of Spergel fitting: a test case exemplification}

Fitting Spergel models to interferometric data can be complex and requires deep high S/N data. However, it can also potentially brings powerful insights and enable investigations into new scientific questions. 

We aim to demonstrate the potential of our approach by re-examining two cases of published ALMA observations of distant sources that were claimed to contain giant halos surrounding the central galaxies (PACS-787 from \citet{silverman18}, including several coauthors of this paper, and HZ7 from \citet{lambert23}). These two specific examples are taken from a growing body of results that report the existence of large halos, often observed in [CII]$\lambda158\mu$m and other tracers \citep[e.g.,][etc]{ginolfi17,fujimoto19,fujimoto20,pizzati20,cicone21,herrera-camus21,jones23,li23,posses23,scholtz23}.  Extended halos around galaxies are generally interpreted as evidence for accretion, outflows, tidal stripping, or other phenomena affecting distant galaxies. This presents a relevant opportunity for further constraining these processes.
However, it is worth considering whether these halos are genuine different structures from the galaxies or simply the outer scale extension of high \sersic-index profiles. In many cases, simple Gaussian fitting was attempted for the central galaxies, and high \sersic\ index profiles are well known to display large halos \citep[e.g.,][]{mancini10}.

We have downloaded the data for both datasets as described below. 
In ALMA Cycle 6, HZ7 was observed with 80 minutes of on-source integration time in the C43-4 configuration with 47/45 12m antennas in band 7 (Project 2018.1.01359.S; PI: M. Aravena). The [CII] emission at 303.93GHz falls into one of four SPWs with a native channel width of 15.625 MHz. 
PACS-787 was observed with high (C40-6, 43 12m antennas with a maximum baseline of 1.1 km) and low (C40-1, 42 12m antennas with a maximum baseline of 278.9 m) resolution configuration in Band 6 with 19.7 and 10.2 minutes of on-source integration time in ALMA Cycle 4 (Project 2016.1.01426.S; PI: J. Silverman). The CO(5-4) emission at 228.17GHz falls into a native channel width of 3.91 MHz in high-resolution observation and 15.62 MHz in low-resolution observation. The calibration targets for the three observations above are J1058+0133 for bandpass, pointing, and flux, and J0948+0022 for phase.

The left panels of Fig.~\ref{fig:real-data} show the dirty images of PACS-787 and HZ7, where emission on large scales of several arcsec (tens of kpc) can be readily seen. We modeled the emission in the $uv$-space with simple Spergel profiles. For PACS-787, which contains two galaxies in the process of merging, we used two Spergel components (one for each galaxy), while for HZ7, we used a single Spergel. The Spergel fitting in both cases is able to fully account for the emission from the galaxies, simultaneously reproducing the inner emission and the outer halos. The residuals are clean, and no further components are needed, as shown in the right panels of Fig.\ref{fig:real-data}. For the case of PACS-787, the two Spergel component have $\nu=-0.36, -0.14$ and $R_{\rm e}/\theta_{\rm b}\sim0.7$ for both. 
For HZ7, we find $\nu=-0.42$ and $R_{\rm e}/\theta_{\rm b}\sim1.3$.
Based on Eq.(\ref{eq:conversion}) and Fig.\ref{fig:index-matching}, this corresponds to \sersic\ $n\sim1.5$--3, which is well above the Gaussian approximation ($n=0.5$) and also above the exponential case ($n=1$), although not as steep as a {\it de\ Vacouleurs} profile ($n=4$). 

We have shown that the full emission in these two systems can be fit with a single component model, which raises questions about the interpretation of the outer emission as a halo. Although high \sersic \ values indicate the presence of both a central component and a profile extending further out than a disk model with both smoothly connected, there is no apparent solution of continuity between the inner parts and the outer halos. This is similar to the inner versus outer parts of elliptical galaxies, and therefore suggestions of different physical origins for them are less substantiated.

We emphasize that we have only re-evaluated two cases of halos from the literature (out of many more existing) to exemplify the power of fitting more complex Spergel models to ALMA data. It is beyond the scope of this work to re-analyse all similar observations from the ALMA archive. However, we obviously anticipate that in other cases, the halos might disappear once fitted with a Spergel model. This is not only because of the analysis presented here but also based on one of the key results from our forthcoming companion paper (Q. Tan et al. in preparation), which suggests that $n>1$ models are required for most ALMA observations of distant galaxies.

\section{Conclusions}\label{sec:conclusion}

The Spergel's Bessel-function-based luminosity profile is a good approximation to \sersic \ profile and has the significant advantage of being analytic with a simple Fourier transform. 
We have performed a thorough analysis of the new Spergel fits method for visibilities in the $uv$-plane, comparing it to the \sersic \ fits for imaged data. Our study aims to assess the effectiveness of the Spergel model fitting based on visibility to galaxy light profiles. We have also tested the robustness of fitting in the $uv$-plane by using simpler forms of point and Gaussian profiles. The main findings of our study are:

\begin{enumerate}
    \item The conversion of Spergel $\nu$ into \sersic \ $n$ can be closely approximated by a two-variable function $n(\frac{R_{\rm e}}{\theta_{\rm b}}, \nu) \sim 0.0249 \frac{R_{\rm e}}{\theta_{\rm b}} {\rm exp}(-7.72 \nu) + 0.191 \nu^2 - 0.721 \nu + 1.32$ in the ($\frac{R_{\rm e}}{\theta_{\rm b}},\nu$)-plane. In most cases, the differences between the best-fit value and the one measured from simulated data for \sersic \ $n$ are found to be within 10\%.
    
    \item When wishing to compare results from Spergel to \sersic\ fitting regarding sizes and total fluxes, similar conversions need to be applied. We find that both the size and total flux estimated by {\it galfit} using a \sersic\ profile tend to be larger than when using Spergel, as the profile becomes steeper than an exponential profile, while the fitted parameters of axis ratios and position angles are unaffected. The variation of both the ratio of $R_{\rm e,se}/R_{\rm e}$ and $S_{\rm se}/S_{\rm sp}$ can be described by a similar two-variable function of $r(\frac{R_{\rm e}}{\theta_{\rm b}}, \nu) \sim p_1 (\frac{R_{\rm e}}{\theta_{\rm b}})^2 {\rm exp}(p_2 \nu + p_3 \frac{R_{\rm e}}{\theta_{\rm b}}) + p_4 \nu + p_5$. The best-fit coefficients for the size ratio of $R_{\rm e,se}/R_{\rm e}$ are $p_1 = 0.00138$, $p_2 = -8.96$, $p_3 = 0.260$, $p_4 = -0.0260$, and $p_5 = 0.996$, and for the flux ratio of $S_{\rm se}/S_{\rm sp}$ are $p_1 = 0.00217$, $p_2 = -7.43$, $p_3 = 0.149$, $p_4 = 0.00942$, and $p_5 = 1.00$, respectively.
    
    \item Our MC simulations have shown that fitting directly in the $uv$-plane (rather than imaging the dataset and fitting in the image plane) leads to more consistent and reliable results. The accuracy of fitted structure parameter estimates obtained from $uv$-plane fits using a Spergel profile is significantly higher, with smaller systematic errors and scatters on the recovery of parameters. In comparison, image-based measurements obtained from {\it galfit} using a \sersic \ model tend to have higher systematic biases and larger uncertainties (worse parameter accuracy by a factor of two). 
     
     \item We have verified the reliability of the parameter uncertainties returned by GILDAS \texttt{uv$\_$fit} modeling. The parameter uncertainties are generally somewhat underestimated, but still correct to better than a factor of two.
     
    \item We recommend to attempt full-flagged Spergel profile fitting only to sources detected with a $S_{\rm tot}/\sigma_{\rm b}$ of at least 50. This is needed for minimal accuracy and reliability of the \sersic \ index (converted from the Spergel index) estimates in the $uv$-plane. The corresponding median value of $\sigma(n)/n$ is 0.36, which we deemed as the minimum threshold for a meaningful and accurate estimate of \sersic \ $n$.
    
    \item The total flux and size estimates obtained from Spergel fitting show larger uncertainties at fixed S/N compared to Gaussian and point profile functions, which have fewer degrees of freedom. For Spergel profile fitting, the uncertainties in measuring galaxy shape parameters were found to be significantly higher than those in measuring flux density. The least accurate constraint, requiring the deepest data, is the Spergel index.
    \item As a test case, we re-analysed literature claims  for discovery of extended halos surrounding distant galaxies. We find that single Spergel models without any extra added halo can fully explain these observations.
    
\end{enumerate}

High-quality interferometric data, such as now routinely obtained from ALMA and NOEMA, allow us to study the morphology of distant galaxies in their submillimeter band emission. This emission primarily arises from thermal dust and molecular gas, which are closely related to star formation. Fitting a Spergel model in the visibility plane is the preferred method for modeling such emission. This should open-up a new window of  investigation to further our general understanding of the evolution of galaxies, in coming years. 
\bigskip

\begin{acknowledgements}
We thank the anonymous referee for constructive suggestions to improve the paper. We are grateful to the IRAM director, Karl Schuster, for the implementation of the Spergel models as part of the GILDAS software package. QT and ED would like to dedicate this paper to the memory of late Prof. Yu Gao, who offered kind support for this work. R.I.P. We wish to thank Daizhong Liu for useful discussions. This work has been partly funded by China Scholarship Council. QT acknowledges support from the NSFC (grant Nos. 12033004, 12003070, 11803090). CGG acknowledges support from CNES. JPe acknowledges support by the French Agence Nationale de la Recherche through the DAOISM grant ANR-21-CE31-0010 and by the Programme National ``Physique et Chimie du Milieu Interstellaire'' (PCMI) of CNRS/INSU with INC/INP, co-funded by CEA and CNES. A.P. acknowledges support by an Anniversary Fellowship at University of Southampton and by STFC through grants ST/T000244/1 and ST/P000541/1.

This work made use of the following Python libraries: \textsc{Astropy}\citep{astropy2022}, \textsc{Numpy}\citep{harris2020}, \textsc{Matplotlib}\citep{hunter2007}, and \textsc{Emcee}\citep{foreman-mackey2013}.
\end{acknowledgements}

%
   \bibliographystyle{aa} 
   \bibliography{reference} 
%
%
%
%


\begin{appendix} %

\section{Matching Spergel profiles with Sersic profiles through mathematical simulations numerically}
\label{app:analytic-matching}

\begin{figure}[tbp]
\centering
\includegraphics[width=\linewidth]{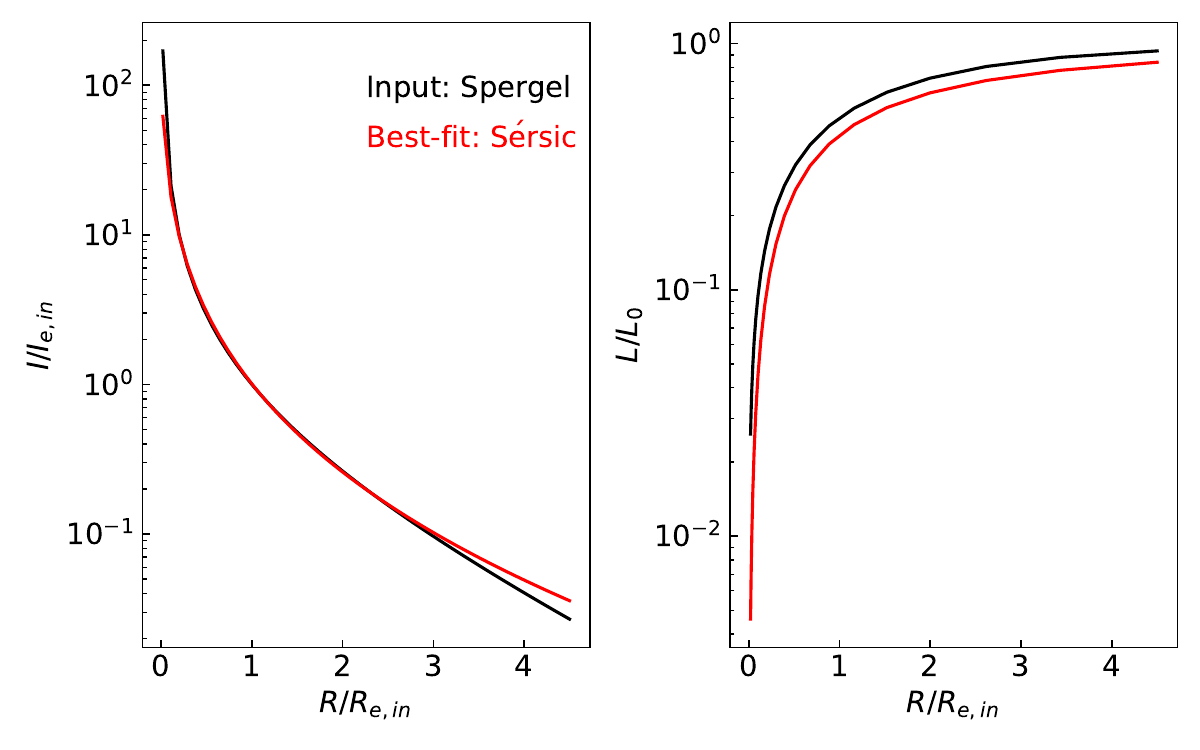}
\caption{\textbf{Left}: surface density profiles for Spergel function (black) compared to the best-fit one with a \sersic \ function (red) through mathematical simulations. \textbf{Right}: comparison of the integrated surface density profiles for the \sersic \ and Spergel functions shown in the left panel. The values of $R_{\rm e,in}$ and $I_{\rm e,in}$ are held fixed and represent the input parameters in the Spergel profile.}\label{fig:app-profile-mcmc}
\end{figure}

Within a valid range of $\nu$ \citep[$-0.85 \leqslant \nu \leqslant 4$; see][]{spergel10}, the matching between Spergel and \sersic \ profiles is derived by minimizing the sum of the difference between the Spergel and \sersic \ functions over a radial range (see Fig.~\ref{fig:profiles} and Fig.~\ref{fig:app-profile-mcmc}), i.e., $\Sigma ({\rm log}((I/I_{\rm e})_{\rm Spergel})-{\rm log}((I/I_{\rm e})_{\rm S\acute{e}rsic}))^2$. We use the \texttt{emcee} package \citep{foreman-mackey2013} in Python to perform Markov chain Monte Carlo (MCMC) search for the best-fit. 

\begin{figure*}[tbp]
\centering
\includegraphics[width=0.3\linewidth]{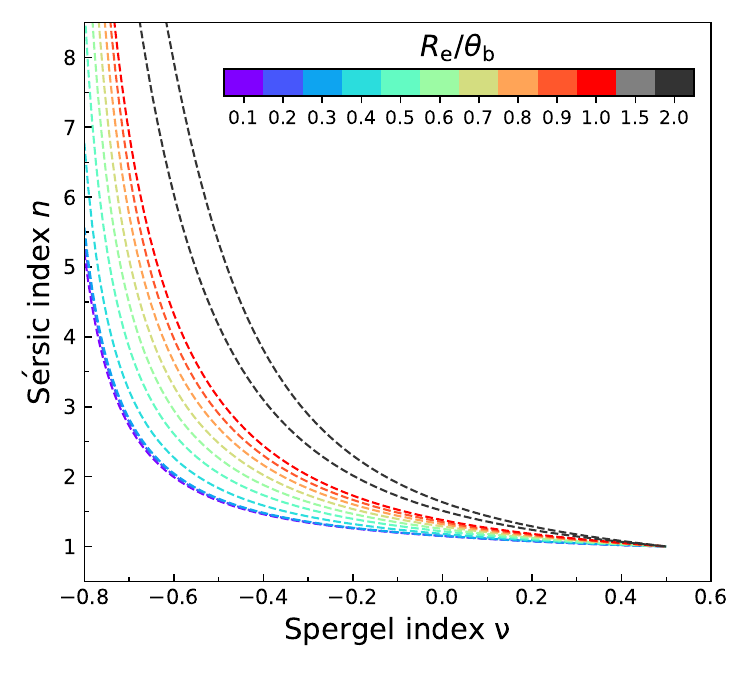}
\includegraphics[width=0.3\linewidth]{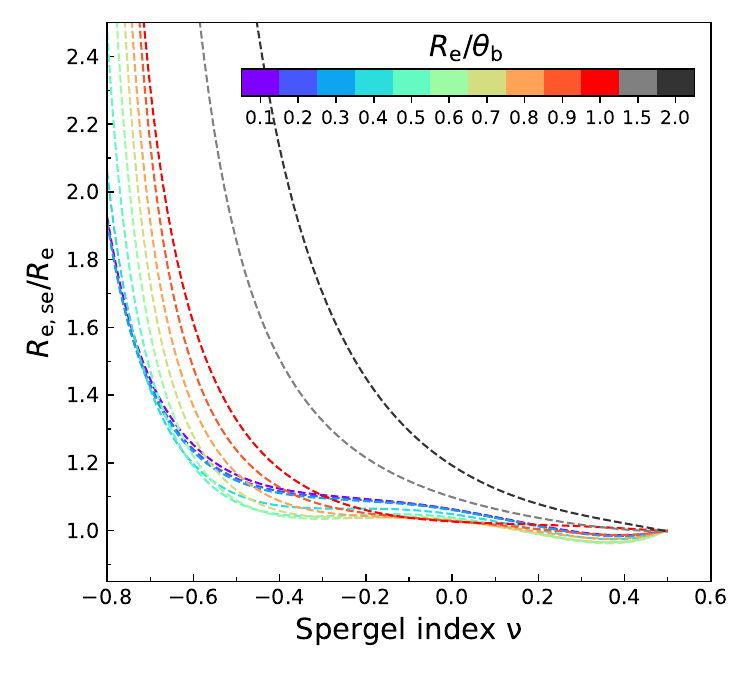}
\includegraphics[width=0.3\linewidth]{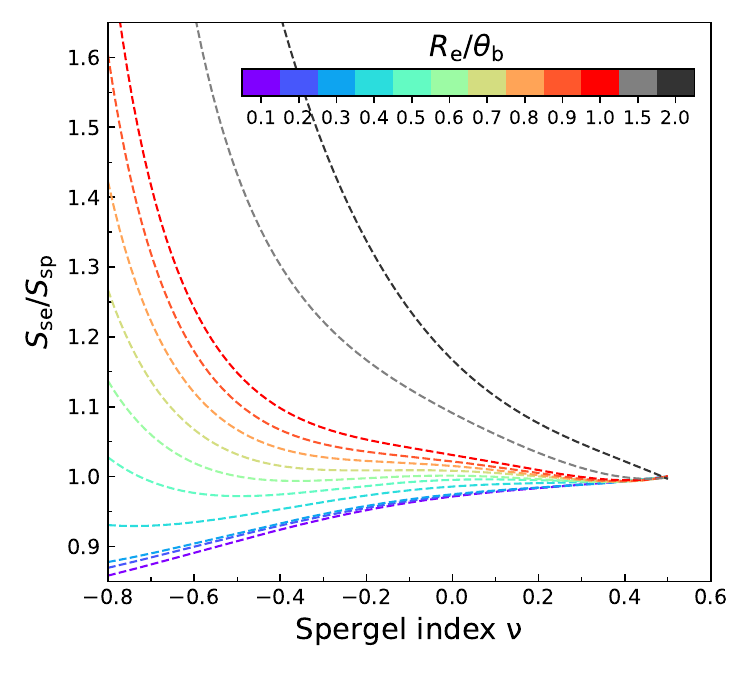}
\caption{Comparison between Spergel index $\nu$ and \sersic \ index $n$ (left), the ratio of $R_{\rm e}$ (middle) and total flux (right) obtained from profile fitting using a \sersic \ function to the input value in the Spergel model as a function of Spergel index $\nu$. }\label{fig:app-match-mcmc}
\end{figure*}

To match with the \sersic \ index measured from {\it galfit} (see Section~\ref{subsec:index-conversion}), we varied the radial range used for analytic matching and found that the best-fit is given by limiting the radius between about 0.01 $\theta_{\rm b}$ and 2.0 $\theta_{\rm b}$ for the source sizes (in units of $\theta_{\rm b}$, $R_{\rm e}/\theta_{\rm b}$) ranging between 0.1 and 2.0. Fig.~\ref{fig:app-match-mcmc} summarizes the output returned from \sersic\ model versus true parameters of the profile fits by setting the intensity, $R_{\rm e}$, and \sersic \ index $n$ as free parameters in the MCMC sampler. We find that the relationship between the \sersic\ output and Spergel input parameters exhibit similar trends as seen in Fig.~\ref{fig:index-matching} and Fig.~\ref{fig:infinite-simu}. 

\section{Details of the three ALMA configurations used to generate simulation data}
\label{app:configurations}

To generate simulation data, three different ALMA array configurations data were used in this work. Table~\ref{tab:configurations} lists the details of each configuration, including the major and minor size of the beam, the PA of the beam, FOV, and the number of antennas.

\begin{table}
\addtolength{\tabcolsep}{-1.5pt}
\caption{Details of the three ALMA array configurations}
\label{tab:configurations}
\centering
\begin{tabular}{lccccc}
\hline\hline
Name & $\theta_{\rm MAJ}$ & $\theta_{\rm MIN}$ & PA & FOV &  N$_{\rm ANT}$ \\
 & (arcsec) & (arcsec) & (degree) & (arcsec) & \\
\hline
Config-A & 0.210 & 0.194 & 34.5 & 18.0  & 40 \\
Config-B & 0.59 & 0.52 & 83.9 & 18.0 & 41 \\
Config-C & 1.10 & 0.85 & 109.3 & 18.8  & 46 \\
\hline
\end{tabular}
\end{table}

\section{Details on the elliptical Spergel profile}
\label{app:spergel:details}

\subsection{Definition and Fourier Transform}

\begin{figure}
  \centering
  \includegraphics[width=\linewidth]{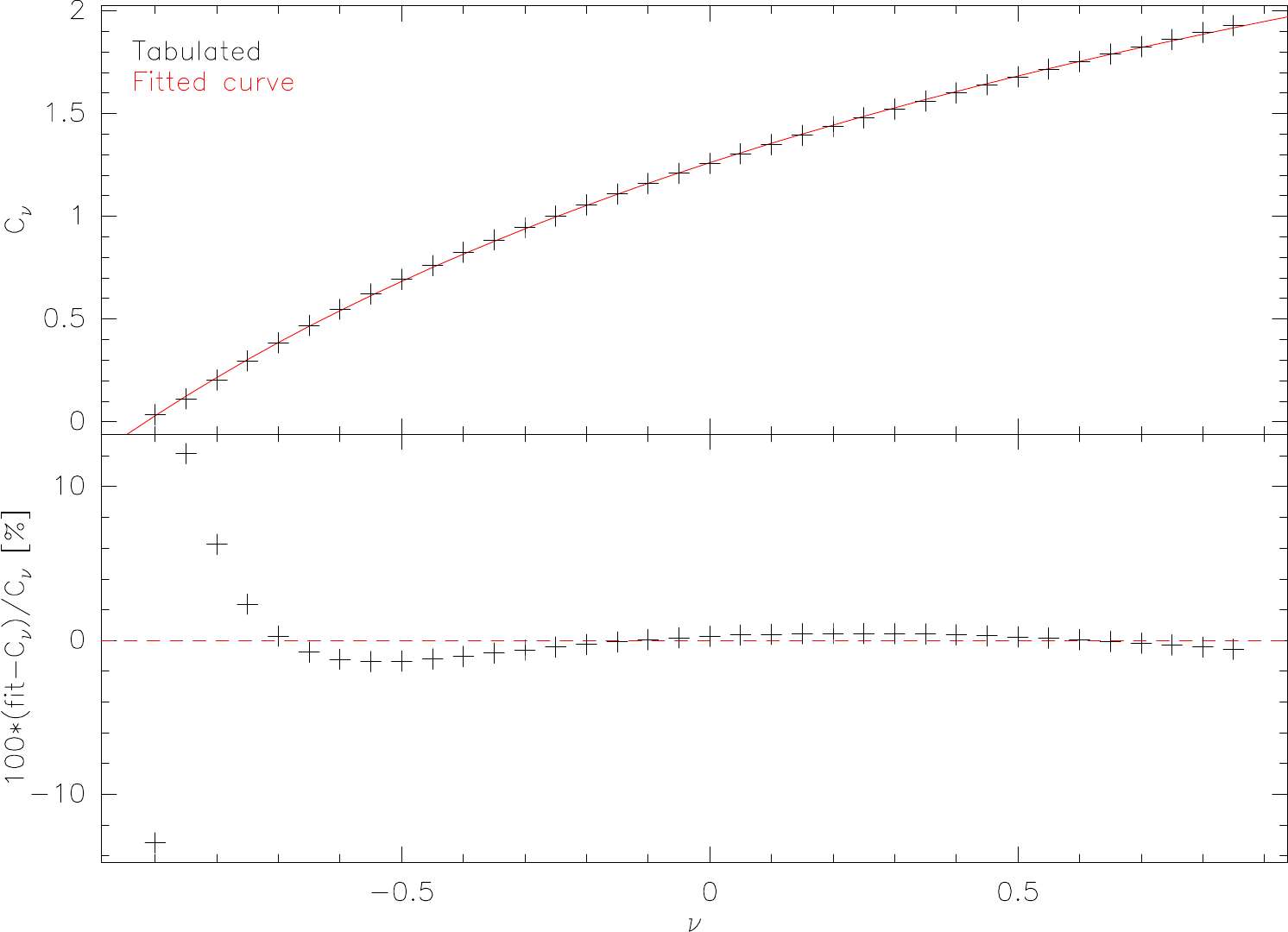}
  \label{fig:cnu}
  \caption{Comparison between tabulated values of the \cnu{} function with
    an analytical approximation.}
\end{figure}

The circular Spergel profile (\spergel{0,0}{circ}) located at the phase center
$(0,0)$ is written in Equation~(\ref{eq:spergel}) as
\begin{equation}
  \spergel{0,0}{circ}(\theta) = \frac{\cnu^2\Lo}{\Ro^2}\,\fnu\paren{\frac{\cnu \theta}{\Ro}},
\end{equation}
where \Lo{} is the total luminosity, \Ro{} is the half light radius,
\cnu{} is a tabulated function of $\nu$ and
\begin{equation}
  \fnu(x) =  \paren{\frac{x}{2}}^{\nu} \frac{\Knu(x)}{\Gamma(\nu+1)},
\end{equation}
in which $\Gamma$ is the Gamma function, and $\Knu(x)$ is the modified
spherical Bessel function of the third kind. \citet{spergel10} tabulates the values of
\cnu{} every $0.05$ for a range of $\nu$ from $-0.90$ to $0.85$. We approximated it with the following approximation
\begin{equation}
  c_{\nu} = \alpha + \beta \nu + \gamma \log(2+\nu)
\end{equation}
with
\begin{eqnarray}
  \alpha & = & -0.403713, \nonumber \\
  \beta  & = & -0.228101, \nonumber \\
  \gamma & = & +2.400961. \nonumber
\end{eqnarray}
Figure~\ref{fig:cnu} compares the tabulated values with our analytical
approximation. It results in relative error of up to 10\% for $\nu < -0.6$
and less than 1\% for $\nu > -0.6$.

We use the radio-astronomy convention to define the conjugate coordinates of
the angular coordinates $(\theta_l,\theta_m)$ relative to the projection
center of the image as $(u,v)$ with
\begin{equation}
  u\,\theta_l= \lambda,
  \quad \mbox{and} \quad
  v\,\theta_m = \lambda,
\end{equation}
where $\lambda$ is the wavelength of the observed line. The
$(\theta_l,\theta_m)$ and $(u,v)$ coordinates are expressed in radian and
meter, respectively. In the $uv-$plane, the Fourier transform of the
circular Spergel profile $\ftspergel{0,0}{circ}(u,v)$ can be written as
\begin{equation}
  \ftspergel{0,0}{circ}(u,v) %
= \Lo \, \bracket{1+\paren{2\pi\,\frac{\Ro}{\cnu}}^2 \paren{u^2+v^2} }^{-(1+\nu)}
\end{equation}
Noting $(\Rmaj,\Rmin)$ the major and minor half light radii, and $\phi$ the
position angle of the elliptical Spergel profile, we yield the following
generalization
\begin{equation}
  \ftspergel{0,0}{elli}(u,v) %
= \Lo \, \bracket{1+\frac{\uvdist}{\cnu^2}}^{-(1+\nu)}
\end{equation}
where $(\urot,\vrot)$ are the coordinates of the $uv-$plane rotated by
$-\phi$ in order to bring the major axis along the $\urot$ axis, i.e.,
\begin{eqnarray}
  \urot &=& u\sin{\PA} + v\cos{\PA},\\
  \vrot &=& u\cos{\PA} - v\sin{\PA},
\end{eqnarray}
and $(\rmaj,\rmin)$ are the reduced major and minor half light radii, i.e.,
\begin{equation}
  \rmaj = 2\pi\,\Rmaj
  \quad \mbox{and} \quad
  \rmin = 2\pi\,\Rmin.
\end{equation}
An additional phase term appears when the Spergel profile is centered at an
offset $(\theta_{l0},\theta_{m0})$ with respect to the phase center.
\begin{equation}
  \ftspergel{\theta_{l0},\theta_{m0}}{elli}(u,v) %
  = \ftspergel{0,0}{elli}(u,v)\,\exp\bracket{2i\,\pi \paren{u\,\theta_{l0} + v\,\theta_{m0}}} %
\end{equation}

\subsection{Partial derivatives}

In order to differenciate the elliptical Spergel profile, we first rewrite it as
\begin{equation}
  \ftspergel{\theta_{l0},\theta_{m0}}{elli}(u,v) = \Lo\,\bracket{\g(u,v)}^{-(1+\nu)}\,\p(u,v)
\end{equation}
with
\begin{equation}
  \g(u,v) = 1+\frac{\uvdist}{\cnu^2},
\end{equation}
and
\begin{equation}
  \p(u,v) = \exp\bracket{2i\,\pi \paren{u\,\theta_{l0} + v\,\theta_{m0}}}.
\end{equation}
The derivative with respect to the luminosity is
\begin{equation}
  \frac{\partial \ftspergel{\theta_{l0},\theta_{m0}}{elli}}{\partial \Lo}(u,v) %
  = \bracket{\g(u,v)}^{-(1+\nu)}\,\p(u,v).
\end{equation}
The derivatives with respect to the offset from the phase center are
\begin{equation}
  \frac{\partial \ftspergel{\theta_{l0},\theta_{m0}}{elli}}{\partial \theta_{l0}}(u,v) = 2i\,\pi\,u\,\ftspergel{\theta_{l0},\theta_{m0}}{elli}(u,v),
\end{equation}
and
\begin{equation}
  \frac{\partial \ftspergel{\theta_{l0},\theta_{m0}}{elli}}{\partial \theta_{m0}}(u,v) = 2i\,\pi\,v\,\ftspergel{\theta_{l0},\theta_{m0}}{elli}(u,v).
\end{equation}
The derivatives with respect to the major and minor half-light radius are
\begin{eqnarray}
  & & \frac{\partial \ftspergel{\theta_{l0},\theta_{m0}}{elli}}{\partial \Rmaj}(u,v) \nonumber\\
  &=& -\Lo \paren{2\pi\,\frac{2\,\rmaj\,\urot^2}{\cnu^2}} \nonumber \\
  & & \times (\nu+1)\,\bracket{g(u,v)}^{-(\nu+2)}\,\p(u,v),
\end{eqnarray}
and
\begin{eqnarray}
  & & \frac{\partial \ftspergel{\theta_{l0},\theta_{m0}}{elli}}{\partial \Rmin}(u,v) \nonumber \\
  &=& -\Lo \paren{2\pi\,\frac{2\,\rmin\,\vrot^2}{\cnu^2}} \nonumber \\
  & & \times (\nu+1)\,\bracket{g(u,v)}^{-(\nu+2)}\,\p(u,v).
\end{eqnarray}
The derivative with respect to the position angle is
\begin{eqnarray}
  & & \frac{\partial \ftspergel{\theta_{l0},\theta_{m0}}{elli}}{\partial \PA}(u,v) \nonumber \\
  &=& -\Lo \bracket{\frac{2(\rmaj^2-\rmin^2)\,\urot\,\vrot}{\cnu^2}} \nonumber \\
  & & \times (\nu+1)\,\bracket{g(u,v)}^{-(\nu+2)}\,\p(u,v).
\end{eqnarray}
Finally, the derivative with respect to the Spergel index is
\begin{eqnarray}
  & & \frac{\partial \ftspergel{\theta_{l0},\theta_{m0}}{elli}}{\partial \nu}(u,v) \\
  &=& \ftspergel{\theta_{l0},\theta_{m0}}{elli}(u,v) \times \nonumber \\
  & & \bracket{%
      (\nu+1)\frac{2(\uvdist)}{\g\,\cnu^3}
      \paren{\beta+\frac{\gamma}{\nu+2}} - \log{\g}}. \nonumber
\end{eqnarray}
\end{appendix}

%
\end{document}